\documentclass[aps,floatfix,twocolumn]{revtex4}

\usepackage{amsmath}
\usepackage{graphicx}
\usepackage{pslatex}
\usepackage{color}
\usepackage[normalem]{ulem}

\newcommand{\sys}{$\mathcal{S}$}
\newcommand{\envone}{$\mathcal{E}_1$ }
\newcommand{\envtwo}{$\mathcal{E}_2$ }

\newcommand{\sysspace}{$\mathcal{S}$ }

\newcommand{\ese}{$\mathcal{E}_1\mathcal{S}\mathcal{E}_2$ }
\newcommand{\esenospace}{$\mathcal{E}_1\mathcal{S}\mathcal{E}_2$}

\begin{document}


\title{Asymmetric temperature equilibration with heat flow from cold to hot in a quantum thermodynamic system}  

\author{Phillip C. Lotshaw}
\thanks{This manuscript has been authored by UT-Battelle, LLC under Contract No. DE-AC05-00OR22725 with the U.S. Department of Energy. The United States Government retains and the publisher, by accepting the article for publication, acknowledges that the United States Government retains a non-exclusive, paid-up, irrevocable, world-wide license to publish or reproduce the published form of this manuscript, or allow others to do so, for United States Government purposes. The Department of Energy will provide public access to these results of federally sponsored research in accordance with the DOE Public Access Plan. (http://energy.gov/downloads/doe-public-access-plan).\\
}
\email{lotshawpc@ornl.gov}
\affiliation{Institute for Fundamental Science, Materials Science Institute, and Department of Chemistry and Biochemistry, University of Oregon \\ Eugene, OR 97403, USA}
\affiliation{Quantum Computational Sciences Group, Oak Ridge National Laboratory \\ Oak Ridge, TN 37830, USA}
\author{Michael E. Kellman}
\email{kellman@uoregon.edu}
\affiliation{Institute for Fundamental Science, Materials Science Institute, and Department of Chemistry and Biochemistry, University of Oregon \\ Eugene, OR 97403, USA}
\date{\today}

\begin{abstract}

A model computational quantum thermodynamic network is constructed with two variable temperature baths coupled by a linker system, with an asymmetry in the coupling of the linker to the two baths.  It is found in computational simulations that the baths come to ``thermal equilibrium" at different bath energies and temperatures.  In a sense, heat is observed to flow from cold to hot. 
A description  is given in which a recently defined quantum entropy $S^Q_{univ}$ for a pure state ``universe"  continues to increase after passing through the classical equilibrium point of equal temperatures,  reaching a maximum at the asymmetric equilibrium.     Thus, a second law account $\Delta S^Q_{univ} \ge 0$ holds for the asymmetric  quantum process.   {In contrast, a von Neumann entropy description fails to uphold the entropy law, with a maximum near when the two temperatures are equal, then a decrease $\Delta S^{vN} < 0$ on the way to the asymmetric equilibrium.}  

\end{abstract}

\maketitle

\section{introduction}

In this paper we investigate a computational model of a multicomponent quantum thermodynamic network in which surprising phenomena are manifest, due to finite-size time-dependent quantum effects.  We observe, in a  straightforward manner of speaking, that by introducing a deliberate asymmetry, heat can be made to flow from cold to hot in a pure state total system consisting of two separate variable temperature baths, coupled through a ``linker system."  {It is important to emphasize more generally that the baths come to a type of equilibrium, in which there is an asymmetry in temperature between the two sides.}  {The heat flow from cold to hot is not a transient fluctuation;  rather it is a spontaneous process in the evolution to a new type of stable asymmetric equilibrium state in quantum thermodynamics.}

We explore the description of these phenomena in terms of a recently introduced \cite{deltasuniv,micro} quantum entropy $ S^Q_{univ}$ for a pure state  ``universe."    {The classically-forbidden process with  final asymmetric equilibrium has a good account in terms of the second law of thermodynamics with the quantum entropy $S^Q_{univ}$, } which gives results in accord with the standard second law  formulation $\Delta S_{univ} \ge 0$.  In contrast, an account in terms of a von Neumann entropy approach, similar to that described by Landau and Lifshitz \cite{LLEntropy} for thermodynamics of large quantum systems,  fails to give a second law entropy account of the equilibration to the unequal temperatures.

 An essential element of our setup is the variable temperature baths.   In a recent paper \cite{variabletame}, we introduced a computational model for such a bath and showed that it comes to thermal equilibrium with a system, while exhibiting quantum thermodynamic effects related to the finite size of the bath.  The variable temperature bath generalized earlier work \cite{polyadbath,deltasuniv,micro,Gemmerarticle,Olshanni,Belgians2} on quantum thermodynamic simulations that used a constant temperature bath.  The body of our work in Refs. \cite{polyadbath,deltasuniv,micro, variabletame} and the current article are built around a largely self-contained exposition in the unpublished dissertation of P.~C.~L., available online \cite{dissertation}.

This work is part of a broad program reexamining the foundations of statistical mechanics in the context of quantum pure states evolving in time \cite{deltasuniv, micro, variabletame, polyadbath, Leitner2015, Leitner2018, Rigol, Deutsch, Deutsch1991, DeutschSupplemental, RigolReview, tasaki1998, Gemmer:2009, Popescu2006, Popescu2009, Goldstein2006, Goldstein2010, Goldstein2015, Reimann2008, Reimann2016, Greiner2016, Porras, Belgians, vNcommentary, vN, vNtrans, PolkovnikovicEntropy, HanEntropy, KakEntropy, Reeb, Sun2014, Gemmerarticle, Wolynes1990, Olshanni, Belgians2, Perez, Rudolph, Partovi}.  There have been other attempts at formulating the second law for quantum pure states \cite{vNtrans,HanEntropy,PolkovnikovicEntropy}, but to our knowledge none of these has yet been associated with new types of quantum thermodynamic behavior such as we have here with $S^Q_{univ}$ in the second law.  An important aspect of $S^Q_{univ}$ is that it can  have ``excess entropy production" $\Delta S^x$ from quantum spreading of the wavepacket \cite{micro} such that $\Delta S^Q_{univ}$ is greater than that expected classically.  We will find that there is a close connection between the anomalous heat flow and $\Delta S^x$ in our time-dependent processes.   

Other recent work has looked at heat flow from cold to hot in different contexts, in theoretical studies of pure states \cite{Rudolph,Partovi} and in mixed states in theory \cite{winter2017} and experiment \cite{Brazilians}.  The mixed state approaches have technical assumptions that make them inapplicable to our pure state approach here.  The most closely related work is a paper of Jennings and Rudolph \cite{Rudolph}, which expanded on the earlier work of Partovi \cite{Partovi}.  Jennings and Rudolph demonstrated how dynamics of entangled  two-level systems could lead to heat flow from cold to hot within subsystems of a pure state.  {Their work predicts transient nonclassical effects like heat flow from cold to hot.  However, they do not discuss new types of quantum equilibrium states, whereas in our work the emphasis is on heat flow from cold to hot along the path to a stable asymmetric temperature equilibrium.  Another important difference is that our initial states are separable, so the asymmetric equilibrium can be accessed from less specialized initial states.  (It should be noted that non-separable initial states could just as well lead to asymmetric equilibrium in our setup.)

\section{Complex Model System with Two Baths} \label{total system}

   \begin{figure}[h]
\begin{center}
\includegraphics[width=5cm,height=5cm,keepaspectratio,angle=0]{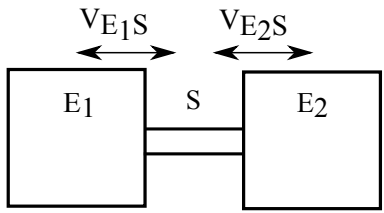}
\end{center}
\caption{Two bath environments $\mathcal{E}_1$ and $\mathcal{E}_2$ are linked by a two-level system $\mathcal{S}$.  The baths exchange heat through the system, with system-environment couplings $\hat V_{\mathcal{E}_1S}$ and $\hat V_{\mathcal{E}_2S}$.} 
\label{setup}
\end{figure}

Fig.~\ref{setup}   shows the setup of  interest:   two variable temperature finite baths or environments  $\mathcal{E}_1$    and $\mathcal{E}_2$ that are uncoupled from each other, except for a system $S$ that acts as a linker.  Each bath is coupled to the linker, but the baths are coupled to each other only indirectly, through the linker.  Suppose the baths start out at different temperatures.  In ordinary classical thermodynamics, the baths and linker system would equilibrate to a common temperature. This would be true even if the couplings to each bath were not  the same. However, we hypothesized  that if we introduce such an asymmetry into a small quantum thermodynamic system, there might be asymmetry of temperature in the final equilibrated state.   This is indeed what we will find.  In this and the  following sections, we describe the setup sketched in Fig.~\ref{setup}, present the results of the computational simulations, and give an account in terms of the quantum entropy  $S^Q_{univ}$.

We consider a linker system  \sysspace with zero-order Hamiltonian $\hat H_\mathcal{S}$ that connects together two finite temperature baths or  environments \envone and \envtwo with Hamiltonians $\hat H_{\mathcal{E}_1}$ and $\hat H_{\mathcal{E}_2}$.  The baths do not interact directly, but rather  interact with the system via coupling operators $\hat V_{\mathcal{E}_1\mathcal{S}}$ and $\hat V_{\mathcal{E}_2\mathcal{S}}$.     The total Hamiltonian is

\begin{equation} \label{Htot} \hat H = \hat H_\mathcal{S} + \hat H_{\mathcal{E}_1} + \hat H_{\mathcal{E}_2} + \hat V_{\mathcal{E}_1\mathcal{S}} + \hat V_{\mathcal{E}_2\mathcal{S}}. \end{equation}

 \noindent The system consists of two levels with energy spacing $\ \hbar\omega_\mathcal{S}=1$ and eigenstates $\{\vert n \rangle\} = \{\vert 0 \rangle, \vert 1 \rangle\}$: 

\begin{equation} \label{HE} \langle n \vert \hat H_\mathcal{S}\vert n \rangle = n. \end{equation}

\noindent The environment (bath) Hamiltonians $\hat H_{\mathcal{E}_1} $ and $\hat H_{\mathcal{E}_2} $ are for identical  collections of $\eta$ harmonic oscillators, each with frequencies $\{\omega_\mathrm{osc}\}$.  The  zero-order eigenstates for bath 1 are $\vert \epsilon_1 \rangle = \vert n_1^{(\epsilon_1)},n_2^{(\epsilon_1)},...,n_\eta^{(\epsilon_1)}\rangle$ with Hamiltonian matrix elements 

\begin{equation}  \langle \epsilon_1 \vert \hat H_{\mathcal{E}_1}\vert \epsilon_1\rangle  = \sum_{osc=1}^\eta \hbar\omega_\mathrm{osc} n_\mathrm{osc}^{(\epsilon_1)}, \end{equation}

\noindent with similar expressions for bath 2 with $\vert \epsilon_2\rangle$ and $\hat H_{\mathcal{E}_2}$.  The frequencies of the bath oscillators are taken as random numbers that are scaled to set their geometric mean $(\prod_{osc=1}^{\eta} \hbar\omega_\mathrm{osc})^{1/\eta}=1$, in accord with Ref.~\cite{variabletame} where this finite environment model was developed as a small variable temperature bath in simulations of a system interacting with a single environment.  The frequencies we use are listed in Table \ref{frequency table}.  We have also observed consistent results in separate calculations using differing sets of frequencies in each bath, so our results do not depend on these specific choices of frequencies or on the use of the same frequencies in each bath.   

\begin{table}[h]
\begin{tabular}{|c|c|c|c|c|c|}
\hline
$\ \eta\ $ & $\hbar\omega_{1}$  &  $\hbar\omega_{2}$  &  $\hbar\omega_{3}$  &  $\hbar\omega_{4}$  &  $\hbar\omega_{5}$    \\
\hline
4 & 0.696\ 987& 0.987\ 490  & 1.088\ 054 & 1.335\ 340 &  \\
 \hline
 5 & 0.620\ 246& 0.735\ 401  & 1.146\ 315 & 1.315\ 886 & 1.453\ 415 \\
 \hline
\end{tabular}
\caption{Oscillator frequencies for the $\eta=4$ and $\eta=5$ oscillator environments, shown to six decimal places.  }
\label{frequency table}
\end{table}

  The interactions $\hat V_{\mathcal{E}_1\mathcal{S}}$ and $\hat V_{\mathcal{E}_2\mathcal{S}}$ in Eq.~\ref{Htot} are selective  random matrix couplings used in Ref.~\cite{variabletame},  similar to those that have long been used in modeling  energy transfer between molecular vibrational modes \cite{Gruebele1995,Gruebele2003}.  For example, $\hat V_{\mathcal{E}_1\mathcal{S}}$ begins with a random matrix $\hat R_{\mathcal{E}_1}$ with elements {$\langle n \vert \langle \epsilon_1 \vert \hat R_{\mathcal{E}_1} \vert \epsilon_1'\rangle \vert n'\rangle = R_{n\epsilon_1,\epsilon_1'n'}$} generated as random numbers from   Gaussian distributions with standard deviation $\sigma=1$.  Each of these matrix elements is then scaled by a coupling constant $k_1$ and by ``taming factors" {$\exp(-\gamma_\mathcal{S} |\Delta n|)$ and $\exp(-\gamma_\mathcal{E} \sum_\mathrm{osc} |\Delta n_\mathrm{osc}^{(\epsilon_1)}|)$ that will be explained shortly.   The final form of the coupling matrix elements is:

\begin{equation}\label{coupling} \langle n \vert \langle \epsilon_1 \vert \hat V_{\mathcal{E}_1\mathcal{S}} \vert \epsilon_1' \rangle \vert n' \rangle = k_1R_{n\epsilon_1,\epsilon_1'n'} e^{-\gamma_\mathcal{S} |\Delta n|}e^{-\gamma_\mathcal{E} \sum_\mathrm{osc} |\Delta n_\mathrm{osc}^{(\epsilon_1)}|}, \end{equation} 

\noindent with a similar expression for $\hat V_{\mathcal{E}_2\mathcal{S}}$.  We set the diagonal elements to zero to preserve the oscillator energies in the zero-order basis, as was done previously in Ref.~\cite{variabletame}, and use only real numbers in the coupling to minimize the numerical overhead.

The coupling constants $k_1$ and $k_2$ set a ``baseline" coupling strength and we want these to exceed a threshold value to obtain good thermalization behavior.  In our previous work of Ref.~\cite{variabletame}, we observed that, for a given total energy $E$ and associated density of states, there exist small $k$ that do not allow for sufficient energy transfer to reach thermal equilibrium between a system and a single bath.  Thus the notion of temperature becomes unclear or undefined.  On the other hand,  good energy transfer and thermalization is observed when $k$ is increased and coupling matrix elements are comparable to or larger than the energy level spacing in the $\mathcal{SE}$ density of states. Here, we choose  minimum $k$ values that ensure good energy equilibration behavior (Fig.~\ref{Equal k bath energies}(a)).  Specific values for  $k_1$ and $k_2$ will vary as described in the results to follow.    The parameters $\gamma_\mathcal{S} = 0.1$ and $\gamma_{\mathcal{E}} = 0.5$ determine how the coupling scales with  quantum  {number differences $|\Delta n|$ and $\sum_\mathrm{osc} |\Delta n_\mathrm{osc}^{(\epsilon_1)}|$ of the coupled system and environment states}.   The larger value for $\gamma_\mathcal{E}$ is needed to obtain physical results such that the environment doesn't spread out too much in energy \cite{variabletame}.  The smaller value of $\gamma_\mathcal{S}$  gives good system thermalization in the dynamical calculations. The parameters $\gamma_\mathcal{S}$ and $\gamma_\mathcal{E}$ have been reduced relative to the values we used for a single system and bath in Ref.~\cite{variabletame}, as we find this to be more successful in obtaining good energy flow between the baths.   

For the basis set we use a ``thermal basis" \cite{micro,variabletame} that is a truncated version of the full tensor product basis $\mathcal{H} = \mathcal{H}_{\mathcal{E}_1}\otimes\mathcal{H}_\mathcal{S}\otimes\mathcal{H}_{\mathcal{E}_2}$.  The thermal basis    contains all basis states in the energy range 

\begin{equation} \label{basis} 0 \leq E_\mathcal{S} + E_{\mathcal{E}_1} + E_{\mathcal{E}_2} \leq E^{\mathrm{max}}. \end{equation}

\noindent   A similar truncated basis was found to give good thermodynamic behavior in Ref.~\cite{variabletame} with a single variable temperature bath of the type we use here. {We find good convergence with $\eta=4$ oscillators using  $E^{\mathrm{max}}=16$ and $E^{\mathrm{max}}=17$ in sections \ref{equal coupling section} and \ref{unequal coupling section} respectively, where the coupling constants $k_1$ and $k_2$ take different values.}  For $\eta=5$ oscillators per bath, we also use $E^{\mathrm{max}}=17$.  With 5 oscillators the calculations are not quite converged.  We are unable to go to higher $E^{\mathrm{max}}$ due to the computational demands of increasing the basis, but the results are qualitatively completely consistent for different $E^{\mathrm{max}}$. 

\section{Temperature and the baths}  \label{temp}

 We will be talking about asymmetries in temperature of two baths, with ``heat flow from cold to hot."  This necessitates a careful consideration of temperature.  Evidently,  we must be  talking about ``temperatures" of the individual baths.  But general formulations of statistical mechanics, and certainly our prior work defining temperature baths \cite{variabletame,micro,deltasuniv,polyadbath}, rely on a notion of a microcanonical ensemble implementation of the thermodynamic temperature 
 
 \begin{equation} \label{T def} T =  (\partial S / \partial E)^{-1}\end{equation}
 
 \noindent for the total system.  In our present case, this would be the two baths + system    $\mathcal{E}_1, \mathcal{E}_2,$ and $\mathcal{S}$.   But this would give a single overall temperature -- not separate, possibly different ``local temperatures" for subsystems i.e. the baths.  The same kind of issue arises when considering ``hot and cold spots" in a larger system with local fluctuations.  Our task is to devise and then implement a sensible idea of ``local temperatures,"  as we now describe. What we are doing is  not unrelated to other ideas of variable temperatures, e.g. the local temperature function in the treatment of non-equilibrium systems of Kondepudi and Prigogine \cite{kp}.  However, an important difference in our work is that we are dealing with a situation with a temperature difference in a state of quantum equilibrium.   (It may help avoid later confusion to point out that in Section \ref{entropy section}, we employ a total quantum entropy $S^Q_{univ}$ -- however, for other purposes than to define a temperature for the total system $\mathcal{E}_1, \mathcal{E}_2,$ $\mathcal{S}$.)

To make progress for the local temperatures, we implement a similar type of thermodynamic temperature definition as for the total system, but without appealing to a microcanonical entropy, which clearly does not apply to the subsystems.  Instead, we use thermodynamics to define sensible temperatures for the individual baths. The essential idea is very simple, with somewhat lengthy details placed in the Appendix.  A brief sketch follows.  To define the  single bath temperatures   we use  the standard thermodynamic relationship applied to  the zero-order levels of the individual baths, for example

\begin{equation} \label{TE def}  \frac{1}{T_{\mathcal{E}_1}} = \frac{\partial S_{\mathcal{E}_1}}{\partial E_{\mathcal{E}_1}}, \end{equation}

\noindent where $S_{\mathcal{E}_1}$ and $ E_{\mathcal{E}_1}$ are the bath entropy and average energy, with a similar expression for second bath temperature $T_{\mathcal{E}_2}$.  We evaluate $T_{\mathcal{E}_1}$ in Eq.~\ref{TE def} analytically as a function of the bath energy, as detailed  in the  Appendix.  The approach uses standard microcanonical relations to calculate subsystem $S_{\mathcal{E}_1}$ and $E_{\mathcal{E}_1}$ for a given total \ese energy.  The temperature $T_{\mathcal{E}_1}$ is then calculated by numerically taking the derivative in Eq.~\ref{TE def}.  This results in a temperature-energy relation $T_{\mathcal{E}_1} = T_{\mathcal{E}_1}(E_{\mathcal{E}_1})$,  with a similar relation for $T_{\mathcal{E}_2}$.  The temperatures increase monotonically with the bath energy  in numerical calculations near our total energy,  with curves similar to the standard type of temperature-energy relationship for an oscillator in an infinite bath. To give an idea of the magnitude of the temperature scale, if we assume that one energy unit equals the 111.77cm$^{-1}$   {vibrational level} spacing of water, then $T =1$ energy unit corresponds to an absolute temperature of $T =160.81$K. The temperatures from Eq.~\ref{TE def} are what we will use throughout our results to compare temperatures in the baths.

   \begin{figure*}
\begin{center}
\includegraphics[width=7.5cm,height=7.5cm,keepaspectratio,angle=0,trim={1.25cm 0 0cm 0},clip]{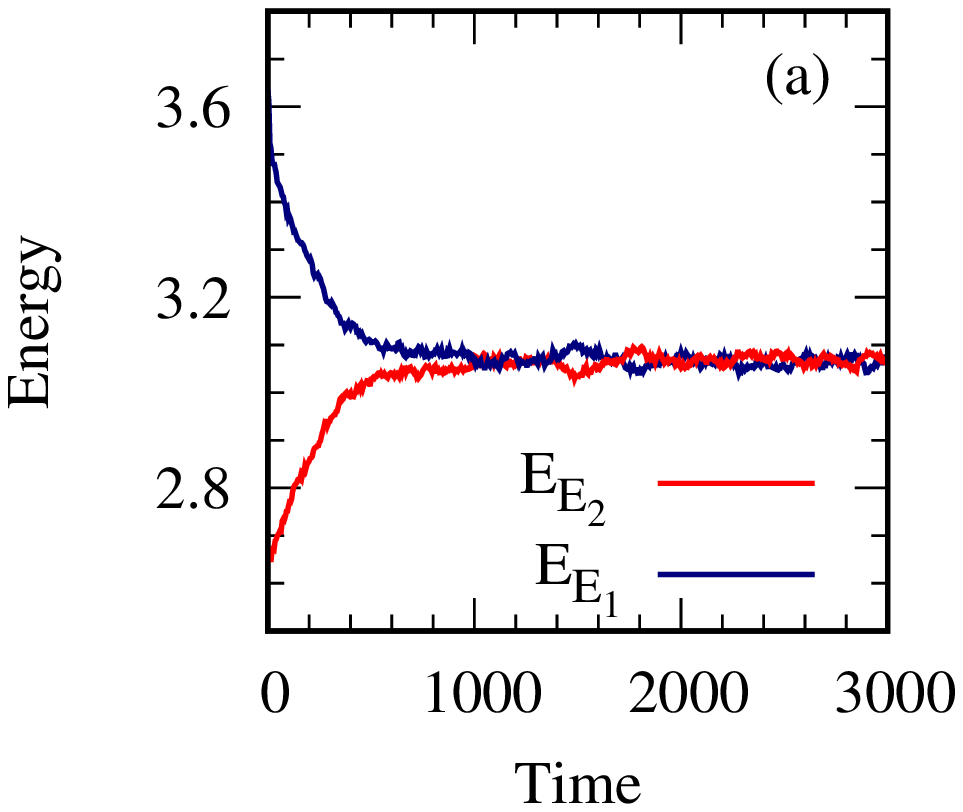}
\includegraphics[width=7.5cm,height=7.5cm,keepaspectratio,angle=0,trim={1.25cm 0 0 0},clip]{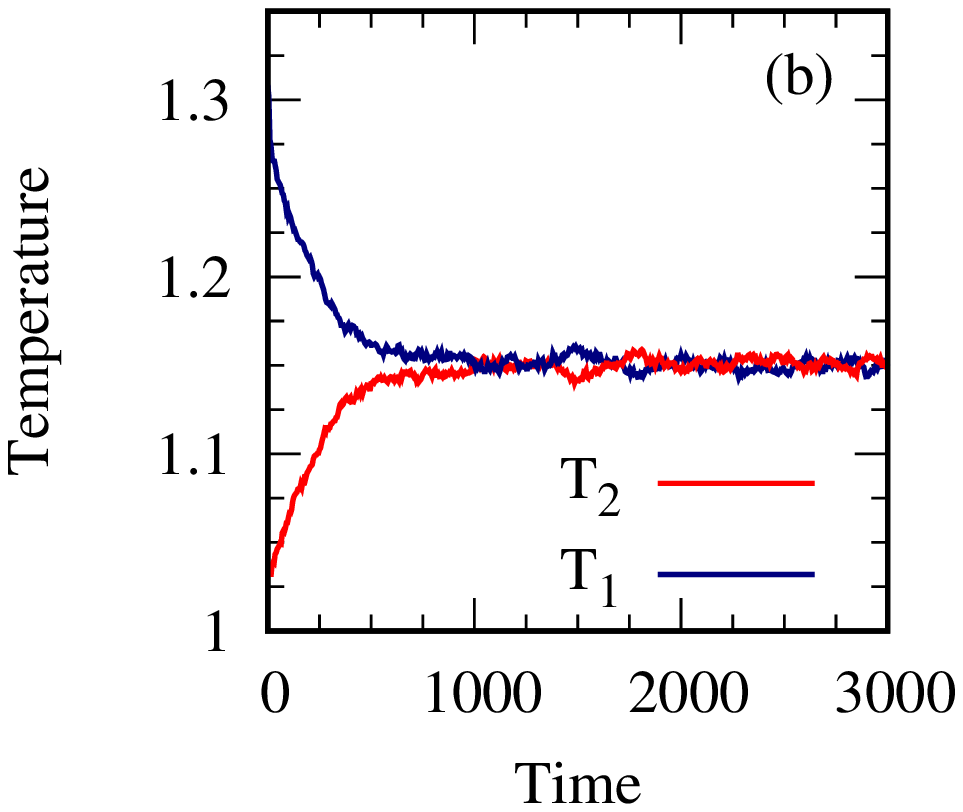}
\end{center}
\caption{Equilibration dynamics with two baths with equal couplings $k_1 = k_2 = 0.085$.  Panel (a) shows the average zero-order energies of the baths, (b) shows the corresponding temperatures from Eq.~\ref{TE def}.} 
\label{Equal k bath energies}
\end{figure*}

\section{Initial States and Time-Propagation}

The initial pure $\mathcal{E}_1\mathcal{SE}_2$ state is a product state:

\begin{equation} \label{initial state} \vert \Psi^0 \rangle = \vert \psi_\mathcal{S}^0 \rangle \vert \psi_{\mathcal{E}_1}^0 \rangle \vert  \psi_{\mathcal{E}_2}^0\rangle \end{equation}

\noindent    The system begins in a single level $\vert \psi_\mathcal{S}^0 \rangle = \vert n \rangle$.  We take the environment states $\vert \psi_{\mathcal{E}_1}^0 \rangle$ and  $\vert  \psi_{\mathcal{E}_2}^0\rangle$ as Gaussian superpositions following Ref.~\cite{variabletame}, for example:

\begin{equation} \label{bath initial state} \vert  \psi_{\mathcal{E}_1}^0 \rangle \sim \sum_{\epsilon_1} \exp(i\delta_{\epsilon_1})\exp(-(E_{\mathcal{E}_1}^0 - E_{\epsilon_1})^2/4\sigma^2) \vert \epsilon_1 \rangle, \end{equation}

\noindent where the $\delta_{\epsilon_1}$ are random phases, $E_{\mathcal{E}_1}^0$ is the central energy of a Gaussian,   $E_{\epsilon_1}$ is the energy of the  zero-order basis state $\vert \epsilon_1\rangle$, and $\sigma = 0.5$ is the width of the Gaussian. A similar expression holds for $\vert \psi_{\mathcal{E}_2}^0\rangle$. We will take different values for $E_{\mathcal{E}_1}^0$ and $E_{\mathcal{E}_2}^0$ in different simulations, as we vary initial energies and temperatures in the baths. A different type of initial bath state with random variations about a Gaussian gave very similar results to those we report here, so our results do not appear to depend significantly on our specific choice of $\vert \psi_{\mathcal{E}_1}^0\rangle$ and $\vert \psi_{\mathcal{E}_2}^0\rangle$ in Eq.~\ref{bath initial state}.

The time-dependent behavior of the total state $\vert \Psi(t) \rangle$ is calculated using a converged Chebyshev polynomial expansion of the time-dependent state.  The expansion is known to give a highly accurate and efficient approximation to the true dynamics.  Detailed accounts of the implementation of the method can be found in Refs.~\cite{Chebyshev, Chebyshev2}.    

\section{Equilibration of the system and baths}

In this section we discuss results of the time propagation for both equal and unequal couplings to the two baths.  In a classical system we would expect the change in couplings to change the rate of approach to equilibrium for the baths, but not their final temperatures,  which we would expect classically to be equal for the two baths.  Does the same temperature independence hold here with the quantum baths with variable couplings?   We will find that this is not the case.   

\subsection{Equilibration with Equal Bath Couplings} \label{equal coupling section}

First we examine the time-dependent behavior with equal coupling constants $k_1 = k_2 =0.085$ in Eq.~\ref{coupling}.  We begin with different initial energies $E_{\mathcal{E}_1}^0 $ and $E_{\mathcal{E}_2}^0 $ in Eqs.~\ref{initial state} and \ref{bath initial state}. The time-dependent behavior of the bath energies is shown in panel (a) of Fig.~\ref{Equal k bath energies}.  The two baths approach an equilibrium state in which  their energies fluctuate about approximately equal values $E_{\mathcal{E}_1} \approx E_{\mathcal{E}_2}$.  In panel (b), we show the time-dependent temperature behavior, based on the single bath temperature $T_{\mathcal{E}}(E_\mathcal{E})$ outlined in Section \ref{temp} and developed in the Appendix.  The temperatures behave very similarly to the energies, ending in an equilibrium state where the two baths fluctuate about the same temperature. This is standard thermodynamic behavior, completely as expected.

\subsection{Equilibration with Unequal Bath Couplings} \label{unequal coupling section}

  \begin{figure*}
\begin{center}
\includegraphics[width=7cm,height=7cm,keepaspectratio,angle=0,trim={1.5cm 0 2cm 0},clip]{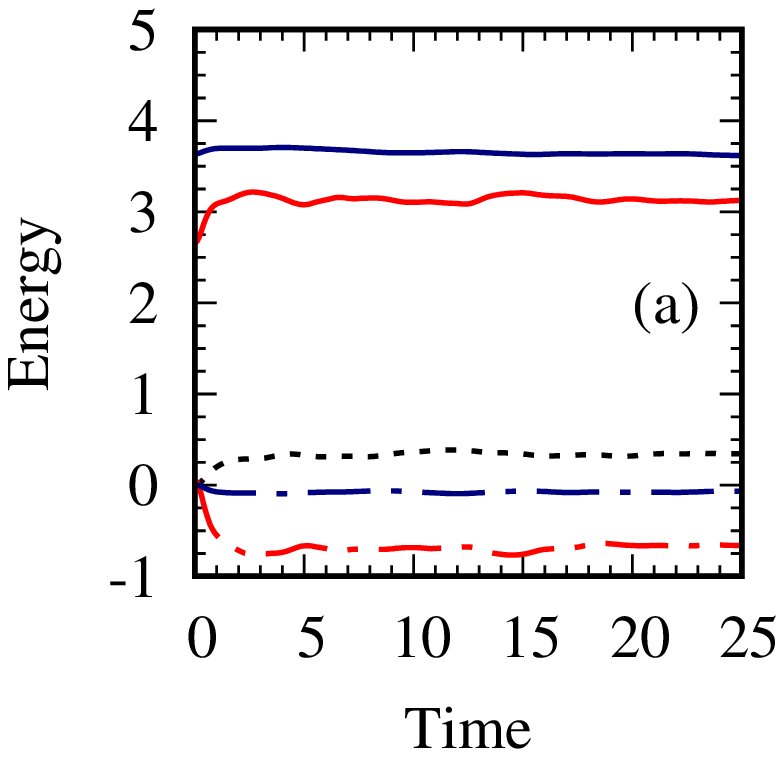}
\includegraphics[width=7cm,height=7cm,keepaspectratio,angle=0,trim={2.5cm 0cm 2cm 0cm},clip]{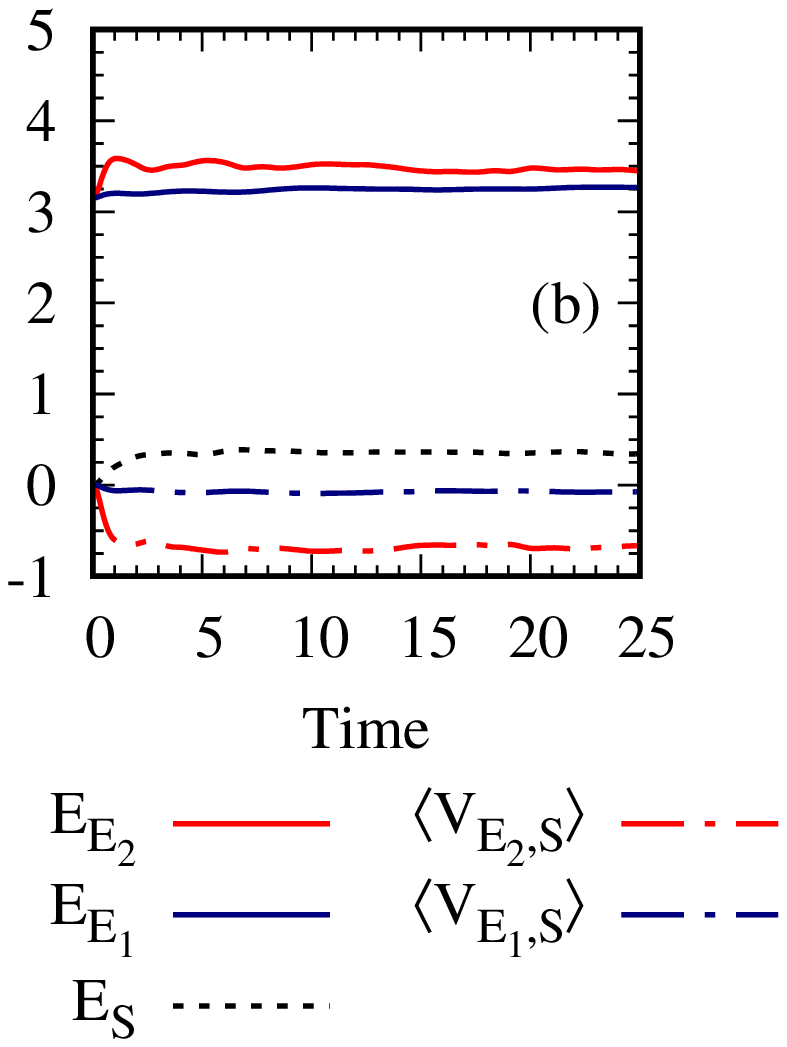}
\includegraphics[width=7cm,height=7cm,keepaspectratio,angle=0,trim={2.5cm 0 2cm 0},clip]{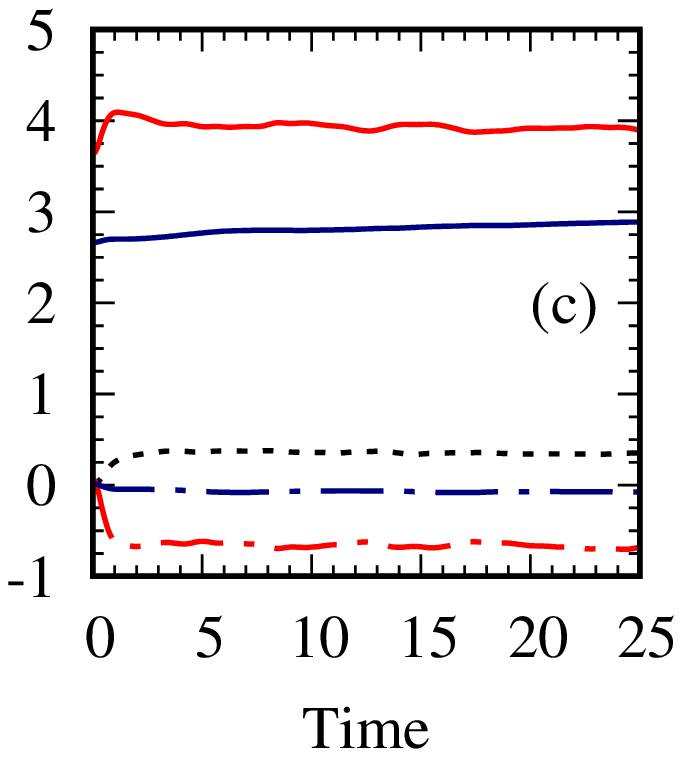}
\end{center}
\caption{Short-time dynamics of energy terms for baths with unequal couplings $k_1 = 0.085$ and $k_2=3k_1$, for different initial bath energies (a) $E_{\mathcal{E}_1}^0 > E_{\mathcal{E}_2}^0$, (b) $E_{\mathcal{E}_1}^0 = E_{\mathcal{E}_2}^0$, and (c) $E_{\mathcal{E}_1}^0 < E_{\mathcal{E}_2}^0$.}
\label{Unequal k bath energies short time}
\end{figure*}

   \begin{figure*}
\begin{center}
\includegraphics[width=6.4cm,height=6.4cm,keepaspectratio,angle=0,trim={1.5cm 0 1cm 0},clip]{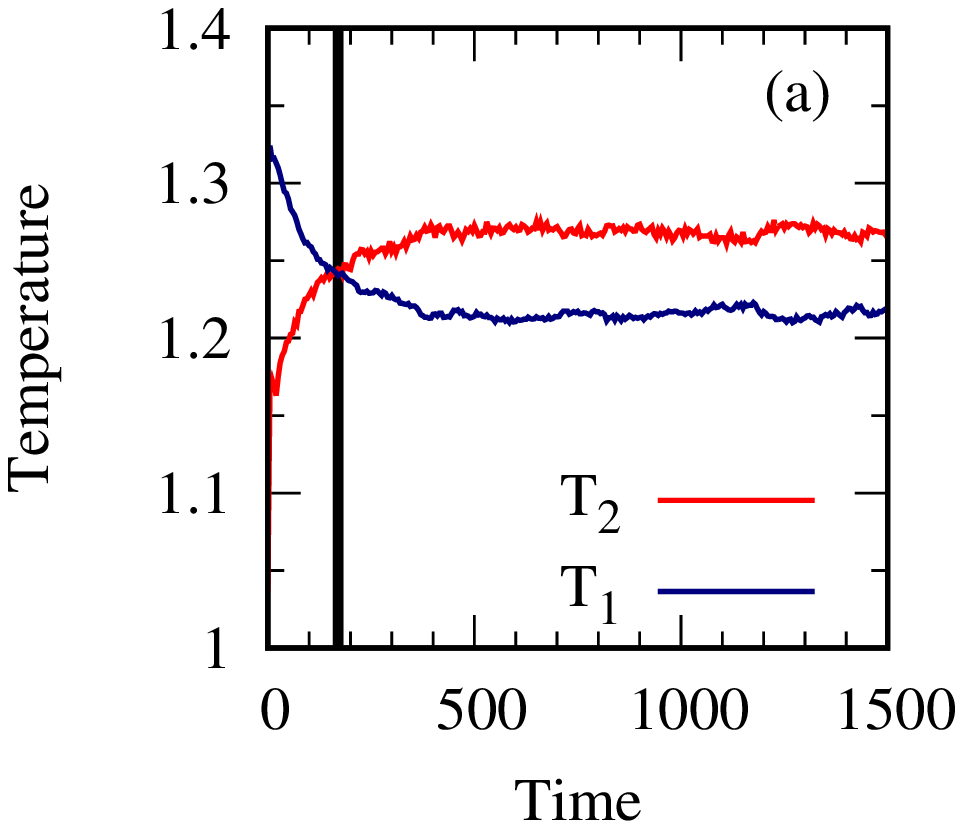}
\includegraphics[width=6.4cm,height=5.6cm,keepaspectratio,angle=0,trim={3cm 0 2cm 0},clip]{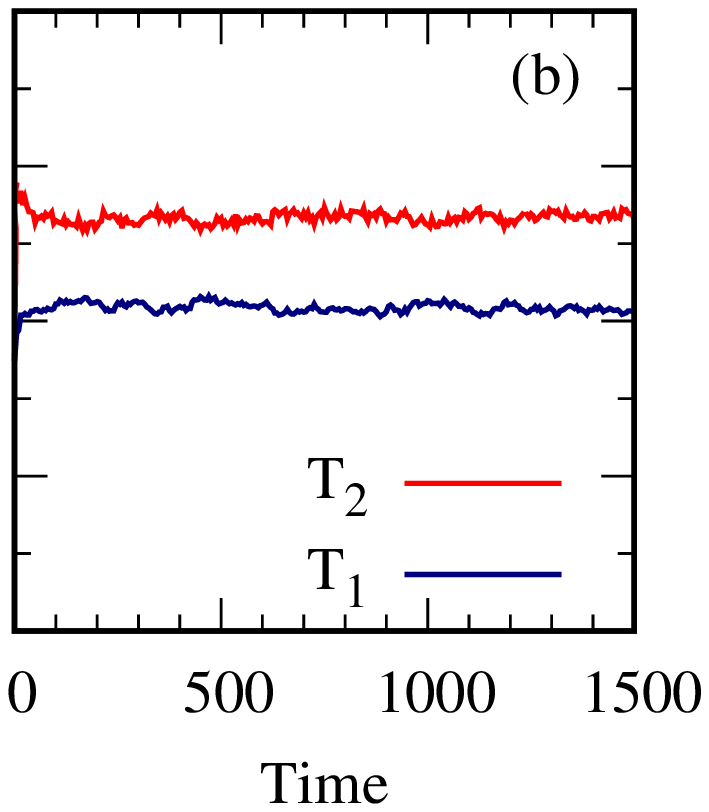}
\includegraphics[width=6.4cm,height=5.6cm,keepaspectratio,angle=0,trim={3cm 0 2cm 0},clip]{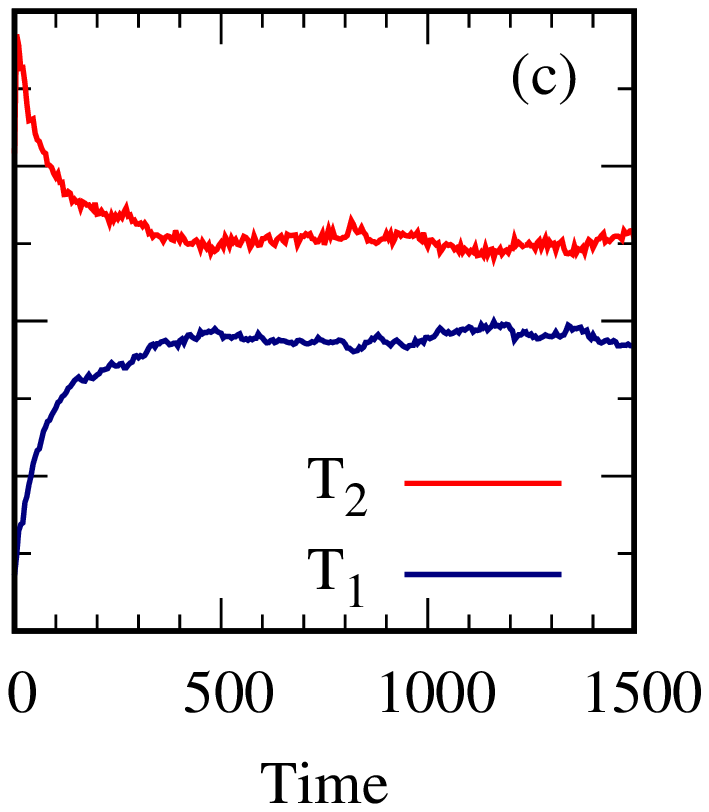}
\end{center}
\caption{Baths with unequal couplings $k_1 = 0.085$ and $k_2=3k_1$ evolve to equilibrium states with unequal temperatures in the baths.  The panels show simulations with different initial bath energies (a) $E_{\mathcal{E}_1}^0 > E_{\mathcal{E}_2}^0$, (b) $E_{\mathcal{E}_1}^0 = E_{\mathcal{E}_2}^0$, and (c) $E_{\mathcal{E}_1}^0 < E_{\mathcal{E}_2}^0$.}
\label{Unequal k bath energies}
\end{figure*}

Next, we examine time-dependent behavior with unequal couplings to the two baths, with $k_1$ = 0.085 and $k_2 = 3k_1$ in Eq.~\ref{coupling}.  To begin, Fig.~\ref{Unequal k bath energies short time} shows details of components of the energy at very short times:  the average coupling energies, bath energies, and the system energy near the beginning of a simulation, for three cases of initial bath energies (a) $E_{\mathcal{E}_1}^0 > E_{\mathcal{E}_2}^0$, (b) $E_{\mathcal{E}_1}^0 = E_{\mathcal{E}_2}^0$, and (c) $E_{\mathcal{E}_1}^0 > E_{\mathcal{E}_2}^0$.  In all cases there is a rapid initial decrease in each of the coupling energies $\langle V_{\mathcal{E}_1,S}\rangle$ and $\langle V_{\mathcal{E}_2,S}\rangle$ with commensurate increases in the bath and system energies.  This behavior will be important when we consider  in Section \ref{why effect} the reasons for the novel effects we are about to describe.

Now we consider our central result: the achievement of an asymmetric temperature equilibration when the couplings to baths are unequal.  In Fig.~\ref{Unequal k bath energies} we show long-time equilibration dynamics of the temperatures $T_{\mathcal{E}_1}$ and $T_{\mathcal{E}_2}$ for simulations with identical parameters to Fig.~\ref{Unequal k bath energies short time}.  The bath-to-bath equilibration happens on the long timescale in Fig.~\ref{Unequal k bath energies} because energy transfer between the baths is indirect and mediated only through the system. Each initial state reaches an asymmetric temperature distribution and remains there, with fluctuations, for all times observed in the calculations.  In panel (a) it is no exaggeration to speak of heat flowing from cold to hot. This is decidedly different from standard thermodynamic behavior!

\section{Entropy}\label{entropy section}

We have seen unusual  behavior in this quantum system:  an equilibrium  in which two temperature baths reach different temperatures, with cases that can be described as having heat flow from cold to hot until an asymmetric equilibrium is reached.  In thermodynamics, we are used to explaining equilibration outcomes with reference to  the second law.  In terms of entropy, this is the statement that $S_{univ}$ reaches a maximum:   {$\Delta S_{univ} = 0$}, given any constraints.    Is anything like this available here?  It might seem not, because quantum statistical mechanics has the von Neumann entropy, and for a pure state, the von Neumann entropy is zero, hence seems to have no relevance.  However, we have recently introduced a new quantum thermodynamic entropy $S^Q_{univ}$ which is nonzero for a pure state \cite{deltasuniv,micro}.  {We generally call this entropy $S^Q_{univ}$ to designate that it is an entropy for the total system-environment ``universe" pure state.}   We have observed in simulations that in ordinary quantum thermodynamic  situations, e.g. heat flow into a single bath, this entropy maximizes at thermal equilibrium, in accord with the second law.  The question naturally arises whether  $S^Q_{univ}$ has salience for the unusual situation of asymmetric temperature equilibration considered here.    We will also try to apply a von Neumann-type entropy analysis using subsystems of the total system, devised with a procedure along  lines discussed by Landau and Lifshitz \cite{LLEntropy}. {We will see that $S^Q_{univ}$ succeeds in giving a second law entropy account of the unusual equilibration.  In contrast, the procedure using the von Neumann entropy fails {for the second law}.}  We  now describe the two approaches, then compare their description of the equilibration process.  

\subsection{Pure state quantum entropy $S^Q_{univ}$}  

The quantum {thermodynamic} entropy $S^Q_{univ}$ was developed in Ref.~\cite{deltasuniv} for an isolated system-environment ``universe" in a pure state $\vert \Psi\rangle$.  It is based in a straightforward way on an expansion of the state in terms of the system-environment zero-order basis  $\vert \Psi\rangle = \sum_{s,\epsilon_1,\epsilon_2} c_{s,\epsilon_1,\epsilon_2} \vert s \rangle \vert \epsilon_1\rangle \vert \epsilon_2\rangle$.  The entropy is then taken along the lines of a Shannon definition using the zero-order probabilities  $p_{s,\epsilon_1,\epsilon_2} = |c_{s,\epsilon_1,\epsilon_2}|^2$ as

\begin{equation} \label{suniv generic} S^Q_{univ} = -\sum_{s,\epsilon_1,\epsilon_2} p_{s,\epsilon_1,\epsilon_2} \ln p_{s,\epsilon_1,\epsilon_2} \end{equation}

\noindent The time evolution of $S^Q_{univ}$ for our simulations, shown in Figs. \ref{entropy fig 4 osc} and \ref{entropy fig 5 osc}, will be discussed along with  calculations in the von Neumann-type approach, to which we turn next.

\subsection{von Neumann entropy approach }

Now we consider an approach based on a von Neumann entropy construct.  Of course, the von Neumann entropy is zero for the total system pure state.  Instead, we follow a procedure similar to that of Landau and Lifshitz \cite{LLEntropy}.  We partition the total system $\mathcal{S} + \mathcal{E}_1 + \mathcal{E}_2$ to define  a von Neumann entropy for each of $\mathcal{S}$, $\mathcal{E}_1$, and $\mathcal{E}_2$.  Specifically, we define the sub-entropies by finding the von Neumann entanglement entropy of each subsystem with respect to the other two subsystems.  Thus $S_\mathcal{S}$ is defined with respect to $\mathcal{E}_1$ and $\mathcal{E}_2$, and so forth for the other combinations.  Then, we find the total entropy    as the sum of the three sub-entropies 

\begin{equation} \label{Svn total} S^{vN}_\mathrm{total} = S_\mathcal{S}^{vN}+S_{\mathcal{E}_1}^{vN}+S_{\mathcal{E}_2}^{vN}   \end{equation}

This procedure seems very reasonable for large total systems, and is likely to give essentially the same, classical results no matter how we divide up the total system.  But in our small quantum system, with the non-classical phenomena in Figs.  \ref{entropy fig 4 osc} and \ref{entropy fig 5 osc}, the behavior of $S^{vN}_\mathrm{total}$  seems {\it a priori} not at all predictable.   

  \begin{figure}[h]
\begin{center}
\includegraphics[width=8cm,height=8cm,keepaspectratio,angle=0]{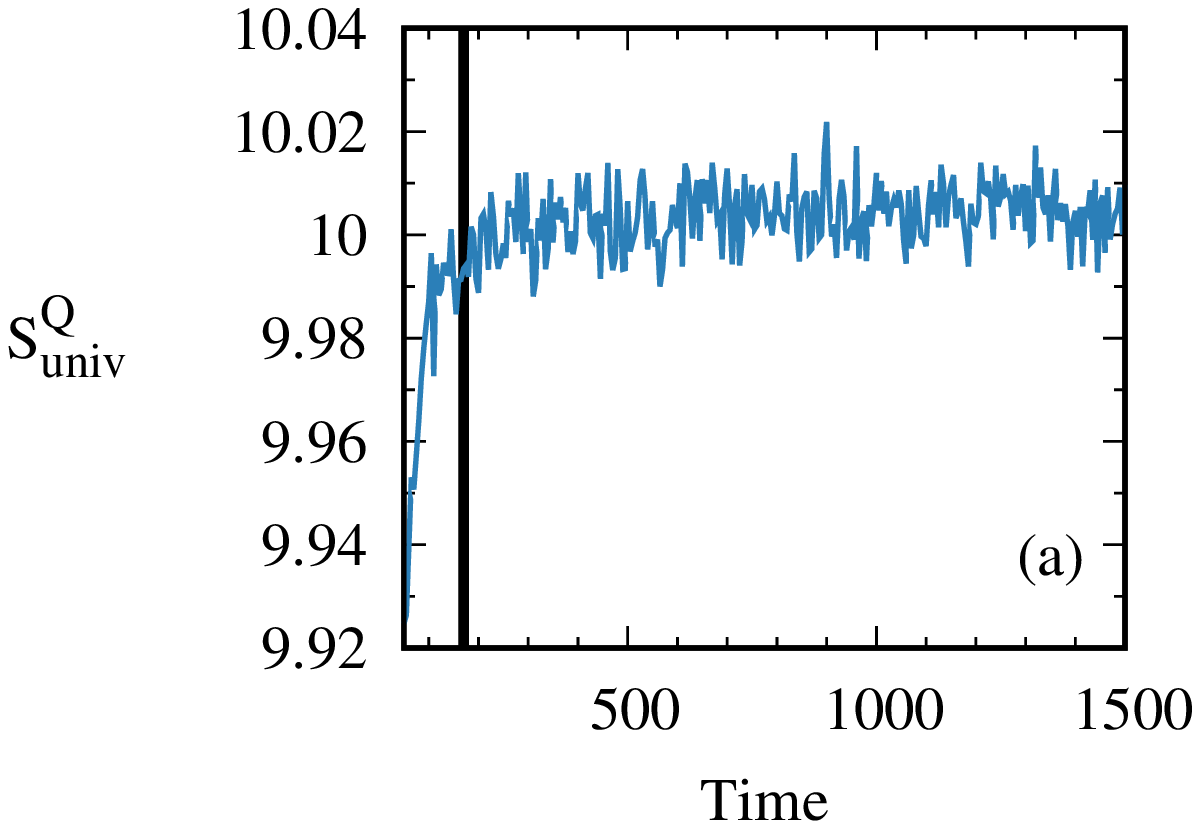}
\includegraphics[width=8cm,height=8cm,keepaspectratio,angle=0]{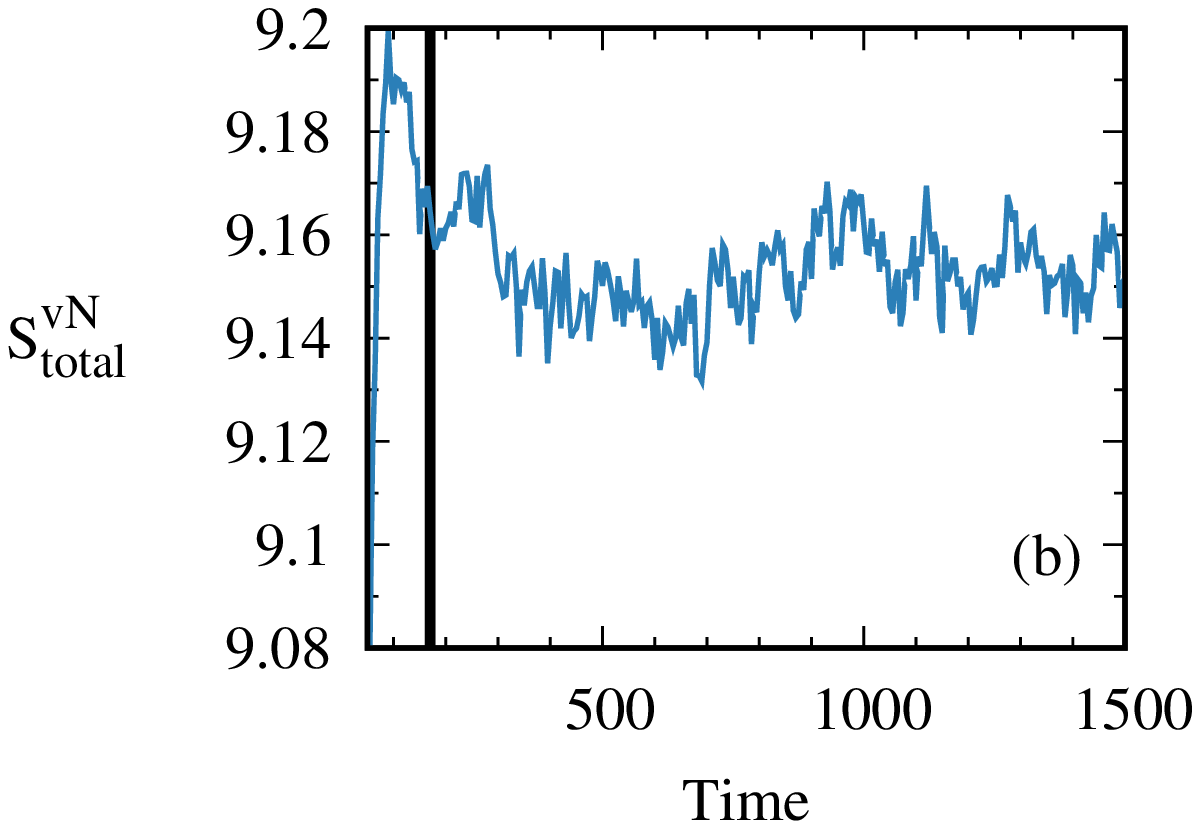}
\end{center}
\caption{(a) Universe entropy production $S^Q_{univ}$ in Eq.~\ref{suniv generic} and (b) $S^{vN}_\mathrm{total}$ in Eq.~\ref{Svn total} for the calculation in Fig.~\ref{Unequal k bath energies}(a)  with $\eta=4$ oscillators per bath.  The black line shows the time of the temperature crossing.  } 
\label{entropy fig 4 osc}
\end{figure}

\begin{figure}[h]
\begin{center}
\includegraphics[width=7.5cm,height=7.5cm,keepaspectratio,angle=0]{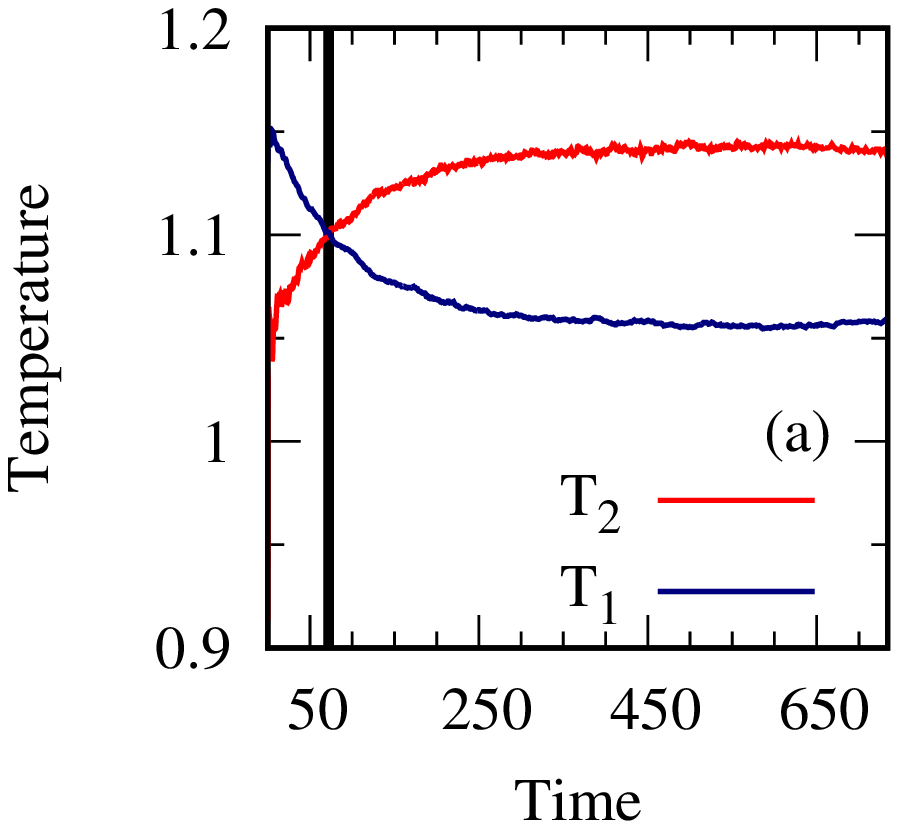}
\includegraphics[width=7.5cm,height=7.5cm,keepaspectratio,angle=0]{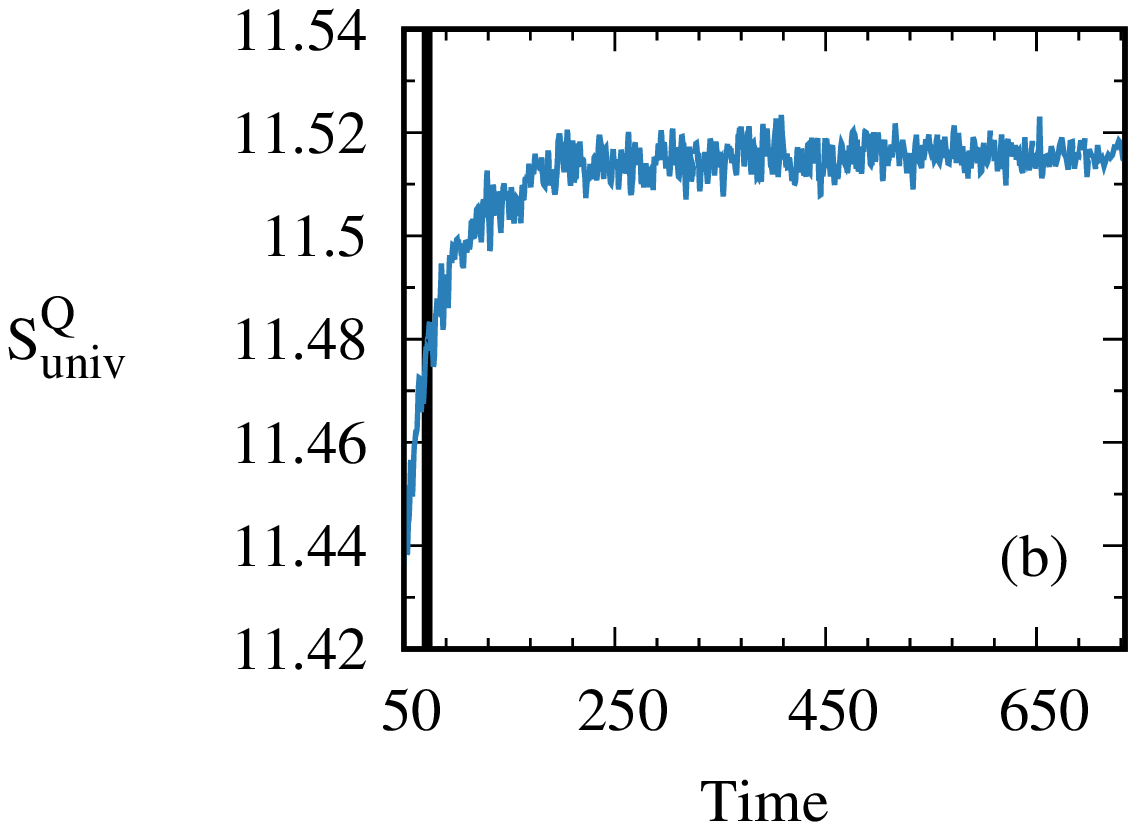}
\includegraphics[width=7.5cm,height=7.5cm,keepaspectratio,angle=0]{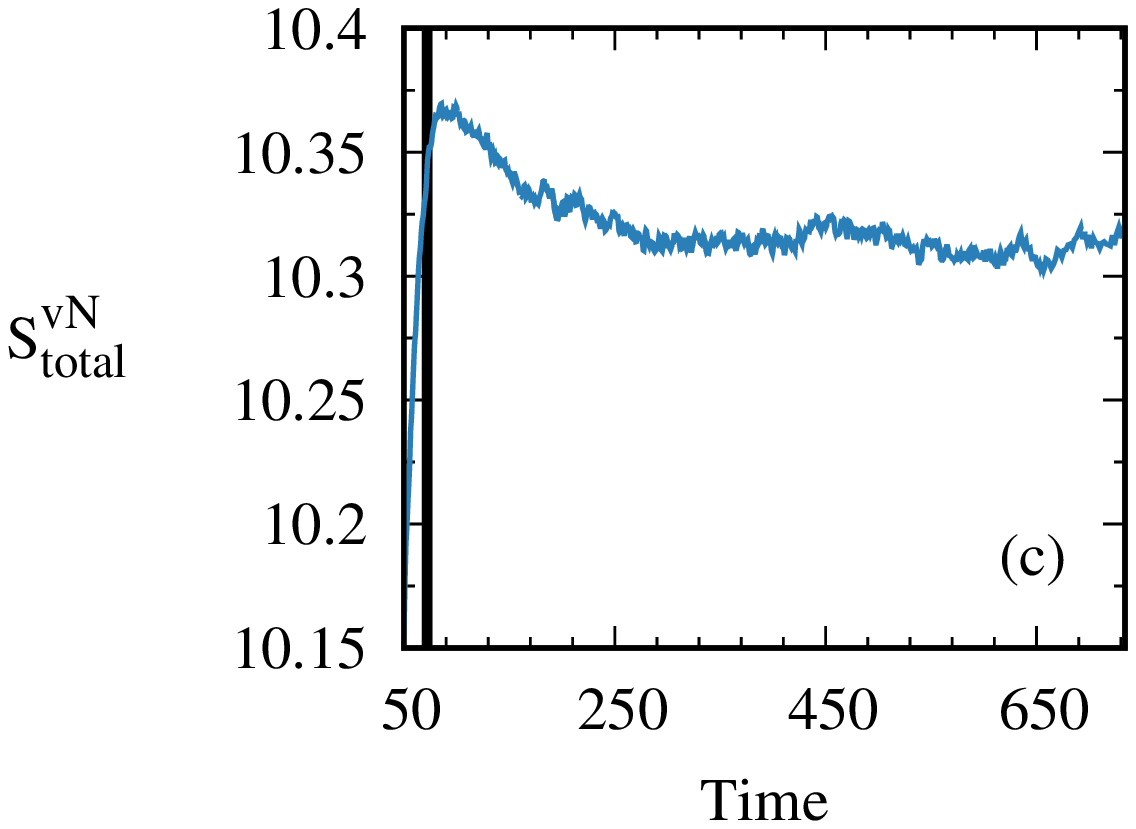}
\end{center}
\caption{
Temperature equilibration similar to Figs. \ref{Unequal k bath energies} and \ref{entropy fig 4 osc} but with $\eta=5$ oscillators per bath.  The increase in $S^Q_{univ}$ and decrease in $S^{vN}_\mathrm{total}$ after the temperature separation is even more striking. } 
\label{entropy fig 5 osc}
\end{figure}

\subsection{Comparison of entropy calculations}  \label{entropycomparisons}

Fig.~\ref{entropy fig 4 osc} shows the two calculations of the total entropy  $S^Q_{univ}$  for the simulation with four oscillators in the top panel of Fig.~\ref{Unequal k bath energies}, with the black line indicating the time of the temperature crossing.   The entropy is calculated as the pure state entropy ${S^Q_{univ}}$ in (a) and as the total von Neumann entropy $S^{vN}_\mathrm{total}$ in (b).

Fig.~\ref{entropy fig 4 osc} shows that $S^Q_{univ}$ rises during the heat transfer process, and is rising still at $T_1 = T_2$. It keeps rising (to within computational fluctuation) until it reaches a maximum at an equilibrium state in which  $T_1 > T_2$.  This shows that $S^Q_{univ}$ is in fact giving a very nice account of  the simulation, consistent with a second law explanation  $\Delta S^Q_{univ} \ge 0$, with $\Delta S^Q_{univ} = 0$ at the nonclassical equilibrium with unequal temperatures.  In contrast, the von Neumann entropy $S^{vN}_\mathrm{total}$   maximizes close to the time of the temperature equalization, then {\it decreases} as the equilibration proceeds to unequal temperatures.  Thus the von Neumann entropy is  maximized near the classical state of equilibrium with equal temperatures, and does not account for the asymmetric quantum equilibrium in terms of entropy and the second law.  

Fig.~\ref{entropy fig 5 osc} shows results where we have increased the number of oscillators per bath to $\eta=5$.  These calculations are not quite converged, as mentioned previously at the end of Section \ref{total system}, but the behavior in Fig.~\ref{entropy fig 5 osc} is robust to changes in the basis size.  Here, we see an even more striking increase in $S^Q_{univ}$  after the temperatures cross, along the way to the asymmetric equilibrium.  On the other hand, the von Neumann entropy is again maximized near the time of the temperature crossing, with a significant decrease as the state approaches the asymmetric equilibrium.  This strongly supports that $S^Q_{univ}$ is giving a correct second law entropy account account of the equilibration behavior, while $S^{vN}_\mathrm{total}$ is not.

It should be noted that a kind of transient  heat flow from cold to hot with a decrease in a sum of subsystem von Neumann entropies has been observed in other work on quantum pure state thermodynamics.  Jennings and Rudolph \cite{Rudolph} developed a model, building on earlier work of Partovi \cite{Partovi}, considering subsystems of a total system in a quantum pure state.  There is no quantum temperature bath in their examples.  Instead, they defined temperature in relation to the reduced density operators of the subsystems. They showed that entangled initial states can have dynamics with heat flow from cold to hot when there is a decrease in the sum of subsystem von Neumann entropies and gave examples for specific correlated initial states of coupled two-level systems.  {A transient spontaneous heat flow from cold to hot  can be seen as an ``ordering" process, and it is perhaps not too surprising that this would be associated with a decrease in a von Neumann measure of the total  entropy.  Certainly, the time-reversed process with heat flow in the normal direction would seem consistent with an increase in this entropy.

The observed decrease in $S^{vN}_\mathrm{total}$ that we see in Figs. \ref{entropy fig 4 osc}-\ref{entropy fig 5 osc} then seems consonant with  the accompanying  heat flow from cold to hot.    But it is important not to exaggerate the parallels between these two lines of inquiry.  Most importantly,    the heat flow from cold to hot of Ref.~\cite{Rudolph} is not associated with an approach to an asymmetric equilibrium as we have here.  Instead it is a transient effect related to specific correlated initial conditions.    In contrast, we begin with uncorrelated separable states.  A variety of different initial states all approach the same type of steady equilibrium state with asymmetric temperatures in the baths.

In sum,  one type of initial state, in Fig.~\ref{Unequal k bath energies}(a),   has heat flow from cold to hot in the approach to the asymmetric temperature  equilibrium.  When this happens there is a decrease in $S^{vN}_\mathrm{total}$.   Thus $S^{vN}_\mathrm{total}$ cannot be used to formulate the second law as  $\Delta S_{univ} >0$.  In contrast, our quantum entropy {$S^Q_{univ}$} gives a successful second law  entropy account of the observed  behavior.   

If we accept this line of thinking, we have still to account for why these nonclassical effects take place starting from our simple separable initial state, and why it is reasonably associated with an {\it increase} in our quantum entropy $\Delta S_{univ} >0$ in apparent accord with a second law formulation.

\section{why the effect takes place} \label{why effect}

We have seen a sizable temperature difference attained in the baths at equilibrium when they have asymmetric couplings, with coupling constants $k_2 > k_1$ in Eq.~\ref{coupling}.  In essence, heat can flow from cold to hot, as seen in Fig.~  \ref{Unequal k bath energies}.    Here we discuss the physical origin of this effect, its relation to the asymmetric coupling, and to ``excess quantum entropy production" in attaining a maximum of $S^Q_{univ}$ at the asymmetric temperature equilibrium.     

We argue that the basic source of these effects in the asymmetric system is that   couplings induce quantum spreading of the wavepacket within the baths during the quantum time evolution.  Interactions among non-resonant energy levels cause the baths to spread to higher and lower zero-order energy basis states $\vert \epsilon_1\rangle$ and $\vert \epsilon_2\rangle$.  Much of the spreading happens on a fast timescale as seen in the dynamics of Fig.~\ref{Unequal k bath energies short time}; {then continues on longer time scales, as reflected in   Fig.~\ref{Unequal k bath energies}. }  
The quantum spreading accesses many more high energy than low energy basis states, since the density of bath states at higher energy is much larger than the density at lower energy, see schematic in Fig. \ref{schematic}.  This asymmetric spreading to mostly high energy basis states has the effect of increasing the energy expectation value of each of the baths and the system, with a compensating decrease in the coupling energies, so that the total energy $\langle \hat H \rangle$ of Eq.~\ref{Htot} is constant.

   \begin{figure}
\begin{center}
\includegraphics[width=7.4cm,height=7.4cm,keepaspectratio,angle=0,trim={1cm 2cm 16cm 2cm},clip]{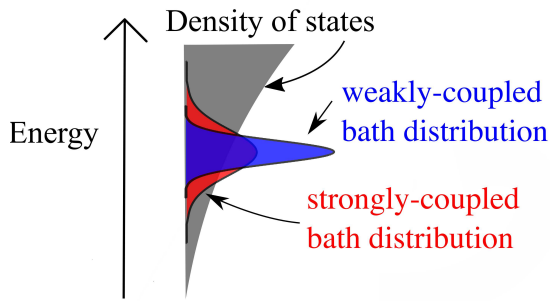}
\end{center}
\caption{Schematic showing why the effect takes place: the strongly-coupled bath distribution spreads to higher and lower energy basis states more than the weakly-coupled bath distribution.  There is a much higher density of states at higher energy, thus the additional spreading favors high-energy states and raises the energy and temperature of the strongly-coupled bath. }
\label{schematic}
\end{figure}

After the initial rapid increase in both bath energies, there is a slower partial transfer of energy from bath to bath via the system, but the energies (and so the temperatures) never equalize in Fig.~\ref{Unequal k bath energies}.  Thus the energy difference from the asymmetric spreading leads to a significant final temperature difference between the baths.  The strongly coupled bath spreads significant probability contributions to a greater number of high energy basis states, so  it ends up at the  higher temperature.   For greater coupling asymmetry we expect greater asymmetry in the spreading and hence a larger temperature difference, while decreasing the coupling asymmetry will decrease the size of the temperature difference, down to equal temperatures when the couplings are equal as in Fig.~\ref{Equal k bath energies}.  In sum,   the asymmetric spreading of the wavepacket to higher energies, along with the incomplete transfer of this energy between the baths, is the basic source of the asymmetric temperature equilibrium. 
 
 Having considered the physical source of the temperature separation, we now consider its connection with   the quantum entropy and the second law.  The entropy $S^Q_{univ}$ of Eq.~\ref{suniv generic} depends on the probabilities $p_{\epsilon_1,s,\epsilon_2}$ of the zero-order basis states $\vert \epsilon_1\rangle\vert s \rangle\vert \epsilon_2\rangle$.
The changes in the $p_{\epsilon_1,s,\epsilon_2}$ and $S^Q_{univ}$ can be thought of qualitatively as a sum of a classical component and a quantum ``excess entropy" component, with $\Delta S^Q_{univ} = \Delta S^{classical} + \Delta S^x$.  The classical $\Delta S^{classical}$ is related to heat flow between the system and baths; alone this would lead to equal temperatures in the baths.  However,  $\Delta S^Q_{univ}$ also depends on the quantum excess entropy production $\Delta S^x$ related to the quantum spreading of the wavepacket to non-resonant energy levels \cite{micro}.  The asymmetric spreading of the wavepacket with asymmetric couplings leads to temperature separation in the baths as described above.  Thus there is a direct connection between $\Delta S_{univ}^Q$, excess entropy $\Delta S^x,$ and the temperature asymmetry of the equilibrium state.  This gives a $\Delta S^Q_{univ}$ that follows the second law $\Delta S > 0$ during the temperature equilibration, as seen in Figs. \ref{entropy fig 4 osc} and \ref{entropy fig 5 osc}.

In contrast, the total von Neumann entropy  $S^{vN}_\mathrm{total}$ of Eq.~\ref{Svn total}  is more indirect  than $S^Q_{univ}$, in that it depends on the eigenvalues of the reduced density operators for the $\mathcal{E}_1, \mathcal{S},$ and $\mathcal{E}_2$ subsystems.  These  evidently do not encode information about the temperature equilibration in such a way that the total von Neumann entropy is maximized at equilibrium,  {as would be}  expected if the standard second law holds in terms of the von Neumann entropy.    Thus, the von Neumann entropy fails to give a second law account of the asymmetric temperature equilibration, in contrast to the quantum entropy $S^Q_{univ}$.}  

 This is all consistent with the preceding considerations on entropy in Section \ref{entropycomparisons}.  To summarize:  The total system is driven toward maximal $S^Q_{univ}$ by the spreading of the time-dependent total state toward an asymmetric equilibrium, in a process of excess entropy production.  The bias comes with the more strongly coupled bath seizing probability density for higher energy states.  This leads to a  {lowering of}  $S^{vN}_\mathrm{total}$ even as $S^Q_{univ}$ is driven to a higher maximum than would happen classically with a symmetric equilibrium.        

We expect the temperature difference to go away or become negligible in the limit of large baths.  The effect is due to energy-uncertainty spreading of the wavepacket to varying energies in the baths, which can be fairly pronounced for small systems as we have here.  However, as the bath size increases the quantum energy uncertainty become less important and ultimately negligible relative to the total energy, so we expect the temperature difference to disappear.}  {On the other hand, it is not yet clear how the temperature asymmetry effect will behave with higher temperatures in the small baths}.  It is not evident to us theoretically that it should disappear, but computational testing is probably beyond present computing capabilities.

It is interesting to consider the source of the asymmetric equilibrium in terms of the eigenstate dynamics in the time-dependent Schr\"odinger equation.  It seems likely to us that the unequal spreading within the baths is encoded into the eigenstates, so that eigenstates usually have temperature separations between the baths.  If so, the temperature separation is likely inevitable at equilibrium, where the coherences between the eigenstates are effectively random and the average behavior of the eigenstates dominates.  This same line of reasoning about the average eigenstate behavior is often cited in the eigenstate thermalization (ETH) hypothesis account of quantum thermodynamics \cite{Deutsch,Rigol}, but to our knowledge this has not yet been extended to novel quantum thermal effects as we have here.  Our hypothesis about temperature separation in the eigenstates appears to us to be entirely consistent with the temperature separation behavior of Fig.~\ref{Unequal k bath energies}, where a variety of initial states all reach approximately the same final temperature-separated equilibrium.   On the other hand, it is not so clear how to relate our results to the ``typicality program" \cite{Popescu2006, Popescu2009, Goldstein2006, vNcommentary, vNtrans, Goldstein2010, Goldstein2015, Reimann2008, Reimann2016} of understanding quantum thermalization and equilibration -- simply because the behavior we are observing is far from what would ordinarily be regarded as ``typical."

\section{summary and prospects}

{In this paper we have considered a quantum total system or ``universe" where surprising thermodynamic behavior is observed: heat flow from cold to hot and a final state of equilibrium with an asymmetric temperature distribution.  This is {certainly}  not standard thermodynamic behavior. 
The anomalous heat flow and temperatures are attributed to quantum effects in the time evolution of two finite environment temperature baths $\mathcal{E}_1$ and $\mathcal{E}_2$ of a few oscillators each, linked together by a system \sys, with \ese collectively in a quantum pure state.   The baths exchange heat indirectly through the system, with system-bath couplings that we varied in strength to examine the resulting temperature equilibration behavior.  By  introducing an asymmetry into this coupling we observe asymmetric temperature equilibrium and anomalous heat flow.}   {The asymmetric temperature equilibrium was explained in terms of asymmetric spreading of the wavefunction in the two baths, with the strongly coupled bath more readily accessing higher energy states than the weakly coupled bath, giving it a higher energy and temperature.  }

Heat flow from cold to hot is never  observed in an isolated classical total system because it decreases the classical entropy of the universe, in conflict with the second law $\Delta S_{univ}>0$.  This raises the question  whether there is a quantum account of the second law that holds for the asymmetric equilibration process, in which the classical account fails.  We examined two different approaches to defining a quantum total entropy $S_{univ}$ to formulate the second law.   The first approach was based on a recently developed entropy $S^Q_{univ}$ for a pure state;  the second used a more conventional type of definition based on a sum of component von Neumann entropies for the system and baths.

{We found that $S^Q_{univ}$ was maximized in the asymmetric temperature equilibrium, with $\Delta S^Q_{univ} > 0$ as heat flowed from cold to hot.  Thus $S^Q_{univ}$ gives a very satisfactory second law account of the observed behavior, with entropy increasing throughout the thermodynamic process.  The success of $S^Q_{univ}$ in describing the temperature separation is related to ``excess entropy production" from spreading of the wavepacket \cite{micro} in the asymmetric equilibrium.  In essence, the spreading of the wavepacket accesses higher energy states in the strongly coupled bath, giving it a higher temperature, and this contributes quantum excess entropy beyond the classical entropy change from heat flow with a fixed total energy.   The second approach to defining a quantum $S_{univ}$ as a sum   $S^{vN}_\mathrm{total}$    of component von Neumann entropies  fails to maximize at the observed equilibrium, instead maximizing around a point of equal temperatures in the baths,  as expected classically.  In sum, we obtain an entirely satisfactory account of the second law with the new  entropy $S^Q_{univ}$, whereas the approach using von Neumann entropies fails.

It is very interesting to consider how the asymmetric temperature equilibration behavior might be observed in experiments or theoretical studies of real pure state total systems.  One possibility is in the dynamics of gas phase molecules, which have recently been considered as promising laboratories for studying fundamental aspects of quantum thermodynamics \cite{Perez,Leitner2015,Leitner2018}.  Pure states of gas phase molecules are relatively easy to prepare and different molecules give access to a wide range of Hamiltonians.  Molecules containing two outer groups of atoms linked together by an atom or small group of atoms could give a total system similar to the \ese we study here, with the outer groups serving as the environments $\mathcal{E}_1$ and $\mathcal{E}_2$ and the linking atom or atoms as the system \sys.  Gas phase experiments or {\it ab initio} calculations \cite{Perez} could look for asymmetric temperature equilibrium in these molecules.  Molecules with similar couplings should be readily available since our coupling is closely related to a generic model for vibrational modes in organic gas-phase molecules.  It could also be very interesting to consider the role of interactions and entanglement of \ese with further exterior systems, for example in interactions with additional molecular degrees of freedom.  If the temperature separation is robust to exterior entanglement and decoherence, then this effect could potentially have technological applications in the condensed phase.   

It is interesting to consider the role of $S^Q_{univ}$ and excess entropy production in more general thermodynamic processes than the simple temperature equilibration process  considered here.  In classical thermodynamics, there can be a variety of thermodynamic variables such as chemical potentials, mole numbers, pressure, and volume that are adjusted together in a thermodynamic process so that the classical entropy $S_{univ}$ is maximized at equilibrium.  It seems entirely  possible that the quantum asymmetries investigated here could extend to these other types of thermodynamic processes as well, with different types of asymmetries in other thermodynamic variables at equilibrium.   If so, $S^Q_{univ}$ with excess entropy production would seem to introduce a great deal of flexibility into quantum thermodynamics beyond what is possible classically.  {In a related vein, it is worth noting that the computational simulations here, and possible experimental realizations e.g. in molecular systems considered above, are for very small systems.  One expects the unusual effects reported here to disappear in the generic large systems limit.  The effect is due to quantum spreading of the wavepacket to states of varying energy, and the magnitude of this quantum energy spreading becomes negligible relative to the total energy $E$ as baths become larger}.  However, if one carefully constructs additional degrees of freedom with their own unusual quantum effects, it is reasonable to anticipate somewhat larger, more complex total systems with interesting non-classical behavior.}

\section*{Acknowledgments} 

We thank Benjam\' in Alem\' an and Dan Steck for stimulating conversations and encouragement.  P.~C.~L. thanks Jeff Cina for interesting discussions of the von Neumann entropy in composite systems, and Rob Yelle and Craig Rasmussen for technical assistance on computations. M.~E.~K. thanks David Perry for many stimulating discussions of quantum thermodynamics.  This work was supported in part by the U.S. Department of Energy Basic Energy Sciences program under Contract  DE-FG02-05ER15634. This work benefited from access to the University of Oregon high performance computers ACISS and Talapas.  This work used the Extreme Science and Engineering Discovery Environment (XSEDE), which is supported by National Science Foundation grant number ACI-1548562.  This work used the XSEDE supercomputer Stampede2 at the Texas Advanced Computing Center through the University of Oregon Campus Champion Allocation TG-TRA170043.  


\section*{Appendix: Temperatures of the individual baths}

{In this section we define and discuss the single bath temperatures we use to analyze temperature equilibration in the main paper.  We begin by considering the standard notion of temperature, based on a total isolated system, in our setup the total system \esenospace.  We then specialize to defining a temperature for the single finite baths $\mathcal{E}_1$ and $\mathcal{E}_2$.  }

Temperature is usually defined with respect to an isolated total system through the fundamental relation $1/T = \partial S / \partial E$ of Eq.~\ref{T def},

\noindent where $E$ is the energy and $S = k \ln W$ is the microcanonical Boltzmann entropy, with $W$ the number of states in the microcanonical ensemble.  For our total system 
\esenospace, this temperature can be evaluated exactly by thinking of the two oscillator baths as a single larger bath and using the equilibrium temperature for a single oscillator bath interacting with a system from Eq.~23  of Ref.~\cite{variabletame}.  However, this temperature is for the total system, whereas we would like to have separate temperatures for the two baths, with potential for the bath temperatures to vary during equilibration or at equilibrium.  For this, we need a notion of temperature that applies to a single bath in our model.

We define the single bath temperatures using the relation Eq.~\ref{T def} applied to the separate baths $\mathcal{E}_1$ and $\mathcal{E}_2$. This yields Eq.~\ref{TE def} of the main text, repeated again for clarity as}

\begin{equation} \label{TE def appendix}  \frac{1}{T_{\mathcal{E}_1}} = \frac{\partial S_{\mathcal{E}_1}}{\partial E_{\mathcal{E}_1}} \end{equation}

\noindent where the bath entropy is

\begin{equation} \label{SE def} S_{\mathcal{E}_1} = -\sum_{\epsilon_1} p_{\epsilon_1} \ln p_{\epsilon_1} \end{equation}

\noindent {and the average bath energy is}

\begin{equation} \label{EE def} E_{\mathcal{E}_1} = \sum_{\epsilon_1} p_{\epsilon_1} E_{\epsilon_1}. \end{equation}

\noindent Both $S_{\mathcal{E}_1}$ and $E_{\mathcal{E}_1}$ are defined in relation to the probabilities $p_{\epsilon_1}$ of the $\mathcal{E}_1$ zero-order microstates $\vert \epsilon_1\rangle$ with energies $E_{\epsilon_1}$ (similar relations hold for the bath $\mathcal{E}_2$ with microstates $\vert \epsilon_2\rangle$ and probabilities $p_{\epsilon_2}$).  To evaluate these expressions, we will derive the $p_{\epsilon_1}$ from the fundamental microcanonical ensemble description of the total system  \esenospace. This gives relations for $S_{\mathcal{E}_1}$ and $E_{\mathcal{E}_1}$ based on standard microcanonical reasoning.  We will use these relations to evaluate the temperature in Eq.~\ref{TE def appendix}, leading ultimately to a temperature-energy relationship $T_\mathcal{E} = T_\mathcal{E}(E_\mathcal{E})$ that we use to calculate the temperature in our simulations.  The details follow. 

{To define the single bath microstate probabilities we begin by considering the fundamental statistical mechanical description of the total \ese system in terms of the microcanonical ensemble at total energy $E$.  In the microcanonical ensemble, each of the \ese microstate trios $\vert \epsilon_1\rangle \vert s \rangle \vert \epsilon_2\rangle$ in the microcanonical energy shell $E - \delta E/2 \leq E_{\epsilon_1} + E_s + E_{\epsilon_2} \leq E + \delta E/2$ is treated as having equal probabilities $p_{\epsilon_1,s,\epsilon_2} = 1/W$, where $W$ is the total number of states in the energy shell.  The probabilities $p_{\epsilon_1}$ for the single bath microstates $\vert \epsilon_1\rangle$ come from adding up probabilities for all of the \ese microstates containing $\vert \epsilon_1\rangle$,   }

\begin{equation} \label{p1 def} p_{\epsilon_1} = \sum_s\sum_{\epsilon_2} p_{\epsilon_1,s,\epsilon_2}, \end{equation}

\noindent (a similar relation holds for the $\mathcal{E}_2$ microstate probabilities $p_{\epsilon_2}$).  We calculate the $p_{\epsilon_1}$ and $p_{\epsilon_2}$ following the method detailed in the next section; in short, we use a continuous density of $\mathcal{E}_2$ states to approximate the discrete sum in Eq.~\ref{p1 def}, leading to continuous approximations for $S_{\mathcal{E}_1}$ and $E_{\mathcal{E}_1}$ from Eqs.~\ref{SE def} and \ref{EE def}.  We use these approximate expressions to numerically calculate the temperature $T_{\mathcal{E}_1}$ in Eq.~\ref{TE def appendix}.  The details of the calculation can be found in the next section; in the remainder of this section we will discuss the behavior of the resulting temperature-energy relationship $T_{\mathcal{E}_1}(E_{\mathcal{E}_1})$.

\begin{figure}[h]
\begin{center}
\includegraphics[width=9cm,height=9cm,keepaspectratio,angle=0]{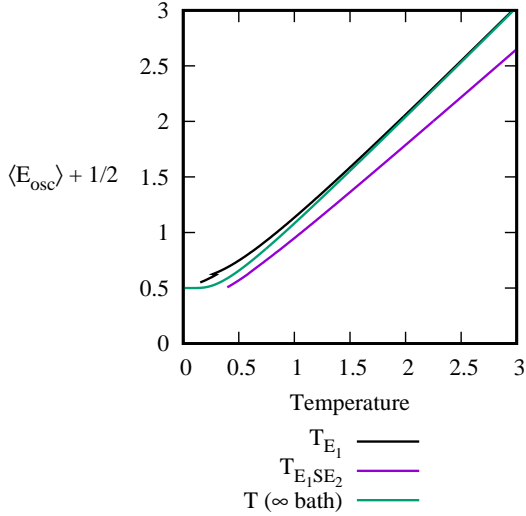}
\end{center}
\caption{Temperature for a single bath $T_{\mathcal{E}_1}$ of Eq.~\ref{TE def appendix} approaches the standard temperature-energy relation for an infinite bath from Eq.~\ref{T infinite bath}.  In contrast, the temperature for the total system $T_{\mathcal{E}_1\mathcal{SE}_2}$ for the total system is higher, as discussed in the text.}
\label{TE fig}
\end{figure}

{Fig.~\ref{TE fig} shows the behavior of the single bath temperature $T_{\mathcal{E}_1}$ of Eq.~\ref{TE def appendix} with $\eta_1=\eta_2=4$ oscillators per bath (the same curve also applies to the temperature $T_{\mathcal{E}_2}$ of the second bath).  $T_{\mathcal{E}_1}$ is compared with the average energy per bath oscillator $\langle E_\mathrm{osc}\rangle +1/2 = E_{\mathcal{E}_1}/\eta_1 + 1/2$ (the factor of 1/2 is an arbitrary added constant that will be explained shortly).  To rationalize the behavior we see in the figure, we will follow a similar route as in our previous work \cite{variabletame} and compare our curve for $T_{\mathcal{E}_1}$ with a more standard type of type of temperature-energy curve from Einstein's 1907 model for the heat capacity of a solid in an infinite temperature bath \cite{EinsteinCollectedPapers1907}.  With an infinite bath, the average number of energy quanta in an oscillator $\langle n_\mathrm{osc}\rangle$ is related to the temperature by}

\begin{equation} \label{T infinite bath} \langle n_\mathrm{osc} \rangle = \frac{1}{e^{1/T} - 1}. \end{equation}

\noindent {We work in units where the energy level spacing of the oscillator is $\ \hbar\omega=1$, so that $\langle n_\mathrm{osc}\rangle = \langle E_\mathrm{osc}\rangle$.  The total energy in the oscillator includes the contribution from energy quanta plus the zero-point energy $\langle E_\mathrm{osc}^{tot}\rangle = \langle E_\mathrm{osc}\rangle + 1/2$, shown along the vertical axis in Fig.~\ref{TE fig}.  The energy starts at the zero-point value of 1/2 at temperature $T=0$, then quickly approaches the equipartition relation $T = \langle E_\mathrm{osc} \rangle +1/2$ at higher energy.  This is the standard behavior with an infinite bath that we will compare with our results for the finite bath temperature $T_{\mathcal{E}_1}$.}

{Now we will analyze the behavior of our single bath temperature $T_{\mathcal{E}_1}$, shown in Fig.~\ref{TE fig} for an example with $\eta_1=\eta_2=4$ oscillators per bath.  For a direct comparison with the infinite bath result, we have plotted $T_{\mathcal{E}_1}$ against $ \langle E_\mathrm{osc} \rangle +1/2$ in the figure, where $\langle E_\mathrm{osc}\rangle =E_{\mathcal{E}_1}/\eta_1$ is the average energy per oscillator.  With the single bath, the factor of 1/2 is an arbitrary added constant needed for a direct comparison with the infinite bath.  The 1/2 does not exactly equal the average zero-point energy, which depends on the variable frequencies of the bath oscillators, see Ref.~\cite{variabletame} for details.  Our temperature $T_{\mathcal{E}_1}$ begins at a non-zero temperature in the figure, where the bath energy is greater than zero.  This is related to the continuous approximation we use, which treats $S_{\mathcal{E}_1}$ as zero in a region around $\langle E_\mathrm{osc} \rangle =0$, where the states are highly discrete and the continuous approximation fails, see the next section for details.  Another note is that there is a discontinuity in the temperature when $\langle E_\mathrm{osc}\rangle +1/2 = 0.625$ in the figure, when the total energy is $E = 1$.  The discontinuity comes from the discontinuous change in the total density of states when the excited system state with energy $E_s = 1$ becomes accessible.  It might be interesting to study discontinuities like this in future studies of thermodynamics of finite systems, but for the purposes of this paper we will focus on higher energies where the temperature is more regular. }

{Now consider the behavior of $T_{\mathcal{E}_1}$ at higher energy, where our continuous approximation is working well.  In this region, $T_{\mathcal{E}_1}$ approaches the standard infinite bath result, where $T_{\mathcal{E}_1} \approx \langle E_\mathrm{osc}\rangle + 1/2$ at high energy.  Thus, at high energy, we have a very normal type of temperature behavior for the single bath temperature $T_{\mathcal{E}_1}$ from Eq.~\ref{TE def appendix}.  The high-energy region is the region we use in our simulations, where we have $T_{\mathcal{E}} \gtrsim 1$. }

{As a final note, we compare $T_{\mathcal{E}_1}$ with the temperature $T_{\mathcal{E}_1\mathcal{SE}_2}$ of the total system.  The total system temperature $T_{\mathcal{E}_1\mathcal{SE}_2}$ is shown by the purple line in Fig.~\ref{TE fig}, and behaves much differently than both $T_{\mathcal{E}_1}$ and the infinite bath temperature.  We compared this type of temperature with the infinite bath temperature in detail in Ref.~\cite{variabletame}, where we showed there is a direct connection between the difference in the temperature curves and the number of oscillators in the bath.  The temperature  $T_{\mathcal{E}_1\mathcal{SE}_2}$ converges to the infinite bath curve as the number of bath oscillators $\eta \to \infty$, as needed in a reasonable temperature definition, with deviations at small $\eta$ corresponding to finite-size effects.  It is interesting to note that in comparison with $T_{\mathcal{E}_1\mathcal{SE}_2}$, the single bath temperature $T_{\mathcal{E}_1}$ is much closer to the infinite bath temperature despite the finite size of the single bath.  The difference is related to energy fluctuations in the single bath - the total system has a fixed energy, whereas $\mathcal{E}_1$ alone has an average energy with significant fluctuations.  Evidently, by the analysis of the figure, these energy fluctuations give a temperature $T_{\mathcal{E}_1}$ that is much closer to the standard temperature with an infinite bath.  

{In summary, we used the standard definition of Eqs.~\ref{T def} and \ref{TE def appendix} to develop temperatures $T_{\mathcal{E}_1}$ and $T_{\mathcal{E}_2}$ for the single baths within the \ese equilibrium state.  The temperatures vary from the temperature of the total system $T_{\mathcal{E}_1\mathcal{SE}_2}$ due to finite size effects and energy fluctuations in the bath.  The final relations $T_{\mathcal{E}_1}(E_{\mathcal{E}_1})$ and $T_{\mathcal{E}_2}(E_{\mathcal{E}_2})$ show approximately standard behavior, close to the standard temperature-energy relation with an infinite bath.  }

\subsection*{Numerical Calculation of the Single Bath Temperature}

In this section we describe our method of numerically calculating the single bath temperature $T_{\mathcal{E}_1}$ from Eq.~\ref{TE def appendix} (similar expressions hold throughout for the second bath $\mathcal{E}_2$ with temperature $T_{\mathcal{E}_2}$).  To calculate $T_{\mathcal{E}_1}$ using Eq.~\ref{TE def appendix}, we need expressions for the single bath entropy $S_{\mathcal{E}_1}$ and average energy $E_{\mathcal{E}_1}$ from Eqs.~\ref{SE def} and \ref{EE def}, which are both defined in relation to the single bath microstate probabilities $p_{\epsilon_1}$ of Eq.~\ref{p1 def}.  Our approach is to approximate the  $p_{\epsilon_1}$ using a density of states function that gives a continuous ``count" of the number of microcanonical states contributing to $p_{\epsilon_1}$.  This leads to tractable continuous expressions $S_{\mathcal{E}_1}$ and $E_{\mathcal{E}_1}$ that we use to numerically evaluate the single bath temperature of Eq.~\ref{TE def appendix} as a converged finite difference.

To begin, consider the expression for $p_{\epsilon_1}$ in Eq.~\ref{p1 def}.  The rightmost sum $\sum_{\epsilon_2}$ counts the number of $\mathcal{E}_2$ states that pair with the $\mathcal{E}_1$ microstate $\vert \epsilon_1\rangle$ and the $\mathcal{S}$ microstate $\vert s \rangle$ in the microcanonical energy shell $E - \delta E/2 \leq E_{\epsilon_1} + E_s + E_{\epsilon_2} \leq E + \delta E/2$, where $E$ is the total energy and $\delta E$ is the width of the energy shell.  For given $\vert \epsilon_1\rangle$ and $\vert s\rangle$, the number of $\mathcal{E}_2$ states $\vert \epsilon_2\rangle$ in the shell can be approximated as $\rho_{\mathcal{E}_2}(E_{\epsilon_2})\delta E$, where

\begin{equation} \rho_{\mathcal{E}_2}(E_{\epsilon_2}) = 
 \begin{cases} 
      \Gamma(\eta_2 + E_{\epsilon_2})/\Gamma(\eta_2)\Gamma(E_{\epsilon_2}+1) & E_{\epsilon_2} \geq 0 \\
      0 & E_{\epsilon_2} < 0\\
   \end{cases}
\end{equation}

\noindent is the density of $\mathcal{E}_2$ states \cite{variabletame} at the central $\mathcal{E}_2$ energy $E_{\epsilon_2} = E - E_s -E_{\epsilon_1}$.  The sum $\sum_{\epsilon_2}$ in Eq.~\ref{p1 def} simply counts the number of $\vert \epsilon_2\rangle$ states, which we approximate as $\rho_{\mathcal{E}_2}(E - E_s - E_{\epsilon_1})\delta E$, giving

\begin{equation} \label{p1 intermediate} p_{\epsilon_1} \approx \sum_s  \rho_{\mathcal{E}_2}(E - E_{\epsilon_1} - E_s)\delta E p_{\epsilon_1,s,\epsilon_2}. \end{equation}

\noindent Next, we consider the microcanonical probability term $p_{\epsilon_1,s,\epsilon_2} = 1/W$, where $W$ is the total number \ese states in the energy shell.  We approximate $W \approx  \rho_{\mathcal{E}_1\mathcal{SE}_2}(E)\delta E$ using the total density of states at the microcanonical energy $E$, 

\begin{equation} \rho_{\mathcal{E}_1\mathcal{SE}_2}(E) = \sum_s \int_0^{E - E_s} dE_{\epsilon_1} \rho_{\mathcal{E}_1}(E_{\epsilon_1})\rho_{\mathcal{E}_2}(E - E_s - E_{\epsilon_1}). \end{equation}

\noindent Putting $p_{\epsilon_1,s,\epsilon_2} \approx 1/ \rho_{\mathcal{E}_1\mathcal{SE}_2}(E)\delta E$ into Eq.~\ref{p1 intermediate}  gives our final expression for $p_{\epsilon_1}$ in terms of the continuous density of states functions

\begin{equation} \label{p1 final} p_{\epsilon_1} \approx \frac{\sum_s \rho_{\mathcal{E}_2}(E - E_{\epsilon_1} - E_s)}{\rho_{\mathcal{E}_1\mathcal{SE}_2}(E)}, \end{equation}

\noindent with a similar expression for the $\mathcal{E}_2$ microstate probabilities $p_{\epsilon_2}$.

With the tractable continuous approximation Eq.~\ref{p1 final} for $p_{\epsilon_1}$, we are now ready to evaluate the single bath entropy and energy of Eqs.~\ref{SE def} and \ref{EE def}.  Putting Eq.~\ref{p1 final} into Eqs.~\ref{SE def} and \ref{EE def} gives 
\begin{equation} \label{SE1 intermediate} S_{\mathcal{E}_1} \approx  -\sum_{\epsilon_1} \left(\frac{\sum_s \rho_{\mathcal{E}_2}(E - E_{\epsilon_1} - E_s)}{\rho_{\mathcal{E}_1\mathcal{SE}_2}(E)}\right) \ln  \left(\frac{\sum_s \rho_{\mathcal{E}_2}(E - E_{\epsilon_1} - E_s)}{\rho_{\mathcal{E}_1\mathcal{SE}_2}(E)}\right)  \end{equation}

\noindent and

\begin{equation} \label{EE1 intermediate} E_{\mathcal{E}_1} \approx \sum_{\epsilon_1} \left(\frac{\sum_s \rho_{\mathcal{E}_2}(E - E_{\epsilon_1} - E_s)}{\rho_{\mathcal{E}_1\mathcal{SE}_2}(E)}\right)E_{\epsilon_1}.\end{equation}

\noindent To further simplify these expressions, we approximate $\sum_{\epsilon_1}$ as an integral, giving the final relations we use to evaluate the single bath energies and entropies in our temperature definition
\begin{align} \label{SE1 final} S_{\mathcal{E}_1} \approx -\int_0^{E} dE_{\epsilon_1} \rho_{\mathcal{E}_1}&(E_{\epsilon_1})  \bigg[\left(\frac{\sum_s \rho_{\mathcal{E}_2}(E - E_{\epsilon_1} - E_s)}{\rho_{\mathcal{E}_1\mathcal{SE}_2}(E)}\right) \nonumber\\
& \times\ln  \left(\frac{\sum_s \rho_{\mathcal{E}_2}(E - E_{\epsilon_1} - E_s)}{\rho_{\mathcal{E}_1\mathcal{SE}_2}(E)}\right)\bigg]  \end{align}

\noindent and

\begin{equation} \label{EE1 final}  E_{\mathcal{E}_1} \approx \int_0^{E} dE_{\epsilon_1} \rho_{\mathcal{E}_1}(E_{\epsilon_1})\left(\frac{\sum_s \rho_{\mathcal{E}_2}(E - E_{\epsilon_1} - E_s)}{\rho_{\mathcal{E}_1\mathcal{SE}_2}(E)}\right)E_{\epsilon_1},\end{equation}

\noindent with similar final expressions for $S_{\mathcal{E}_2}$ and $E_{\mathcal{E}_2}$.  The integrands of Eqs.~\ref{SE1 final} and \ref{EE1 final} have additional factors of $\rho_{\mathcal{E}_1}(E_{\epsilon_1})$ in comparison to the summands from Eqs.~\ref{SE1 intermediate} and \ref{EE1 intermediate}  to account for the fact that there are $\rho_{\mathcal{E}_1}(E_{\epsilon_1})dE_{\epsilon_1}$ summands in each energy interval $dE_{\epsilon_1}$ of integration.  The continuous approximation for $S_{\mathcal{E}_1}$ in Eq.~\ref{SE1 final} fails at very small total energies $E$, where the approximate $S_{\mathcal{E}_1}$ can become negative, whereas the true entropy is strictly non-negative.  We simply take $S_{\mathcal{E}_1} = 0$ in this region and do not evaluate temperatures until $S_{\mathcal{E}_1} >0$. This is the reason for the non-zero minimum energy $\langle E_\mathrm{osc}\rangle$ in the temperature curve of Fig.~\ref{TE fig} of this document.

We are now ready to evaluate the single bath temperature of $T_{\mathcal{E}_1}$ of Eq.~\ref{TE def appendix} using the tractable expressions for $S_{\mathcal{E}_1}$ and $E_{\mathcal{E}_1}$ in Eqs.~\ref{SE1 final} and \ref{EE1 final}.  The temperature $T_{\mathcal{E}_1}$ is calculated numerically in {\it Mathematica} using a finite difference

\begin{equation} \label{TE numeric} \frac{1}{T_{\mathcal{E}_1}} \approx \frac{\Delta S_{\mathcal{E}_1}}{\Delta E_{\mathcal{E}_1}}, \end{equation}

\noindent where $\Delta S_{\mathcal{E}_1}$ and $\Delta E_{\mathcal{E}_1}$ are taken as the differences in $S_{\mathcal{E}_1}$ and $E_{\mathcal{E}_1}$  between two microcanonical states with total \ese energies $E$ and $E + \Delta E$.  We find converged results with $\Delta E = 10^{-6}$, so that the finite difference is an essentially exact approximation to the true derivative of Eq.~\ref{TE def appendix}.  The relation Eq.~\ref{TE numeric}, with $S_{\mathcal{E}_1}$ and $E_{\mathcal{E}_1}$ from Eqs.~\ref{SE1 final} and \ref{EE1 final}, is the final expression we use for $T_{\mathcal{E}_1}$ in Fig.~\ref{TE fig} of this document and in the results of the main paper, with a similar expression for temperature of the second bath $T_{\mathcal{E}_2}$.

\bibliography{twobath}

\end{document}